\documentclass[useAMS,usenatbib]{mn2e}
\usepackage{graphicx,amssymb,natbib}
\usepackage{color}

\title{Star formation in young star cluster NGC 1893}
\author[Sharma et al.]
       {Saurabh Sharma$^1$ \thanks{saurabh@aries.ernet.in}, A. K. Pandey$^{2,1}$, D. K. Ojha$^3$, W. P. Chen$^2$, S. K. Ghosh$^3$,
\newauthor B. C. Bhatt$^4$, Maheswar, G.$^1$ \& Ram Sagar$^1$ \\\\
	$^1$Aryabhatta Research Institute of Observational Sciences (ARIES), Nainital, 263 129, India\\
$^2$ Institute of Astronomy, National Central University, Chung-Li 32054, Taiwan\\
$^3$Tata Institute of Fundamental Research, Mumbai (Bombay) - 400 005, India\\
$^4$CREST, Indian Institute of Astrophysics, Hosakote 562 114, India}
\date{Accepted ......
      Received .....}

\pagerange{\pageref{firstpage}--\pageref{lastpage}}
\pubyear{2007}

\begin{document}

\maketitle

\label{firstpage}

\begin{abstract}

We present a comprehensive multi-wavelength study of the star-forming region NGC 1893 to explore the effects of massive stars on low-mass star formation. Using near-infrared colours, slitless spectroscopy and narrow-band $H\alpha$ photometry in the cluster region we have identified candidate young stellar objects (YSOs) distributed in a pattern from the cluster to one of the nearby nebulae Sim 129. The $V, (V-I)$ colour-magnitude diagram of the YSOs indicates that majority of these objects have ages between 1 to 5 Myr. The spread in the ages of the YSOs may indicate a non-coeval star formation in the cluster. 
The slope of the KLF for the cluster is estimated to be $0.34\pm0.07$, which agrees well with the average value ($\sim 0.4$) reported for young clusters.
For the entire observed mass range $0.6 < M/M_\odot \le 17.7$ the value of 
the slope of the initial mass function, $`\Gamma$', comes out to be $-1.27\pm0.08$, 
which is in agreement with the Salpeter value of -1.35 in the solar neighborhood. 
However, the value of $`\Gamma$' for PMS phase stars (mass range $0.6 < M/M_\odot \le 2.0$) is found to be $-0.88\pm0.09$ which is shallower than the value
($-1.71\pm0.20$) obtained for MS stars having mass range $2.5 < M/M_\odot \le 17.7$ indicating a break in the slope of the mass function at $\sim 2 M_\odot$. Estimated $`\Gamma$' values indicate an effect of mass segregation for main-sequence stars, in the sense that massive stars are preferentially located towards the cluster center. The estimated dynamical evolution time is found to be greater than the age of the cluster, therefore the observed mass segregation in the cluster may be the imprint of the star formation process. There is evidence for triggered star formation in the region, which seems to govern initial morphology of the cluster. 

\end{abstract}


\section{INTRODUCTION}

Young open clusters provide important information relating to star formation process 
and stellar evolution, because such clusters contain massive stars as well as low mass
pre-main sequence (PMS) stars. The age spread in a young open cluster represents the 
cluster formation timescale which can be studied from the analysis of the colour-magnitude diagrams
of very young open clusters. 
To understand the star formation process, it is necessary to know how the star formation proceeds in star clusters and whether the stellar mass distribution that arises from the fragmentation of molecular clouds is universal or 
depends on the local environments. Among various tools, the initial mass function (IMF) is one of the most important tools to study the above mentioned questions (cf. Pandey et al. 2005 and references therein).

An important study to explore the star formation history and IMF of young open clusters was by Phelps \& Janes (1994), who observed 23 open clusters in the Cassiopeia region and estimated IMFs for eight young clusters (Phelps \& Janes 1993). The slope of the upper mass part of the IMF was also systematically investigated by Massey and collaborators using the UBV CCD photometry and MK classification (Massey et al. 1995a; Massey et al. 1995b). They determined the IMF for various associations in the Galaxy and the Magellanic Clouds in a homogeneous manner and found no statistically significant difference in IMF slopes, with an average value of $\Gamma = -1.1\pm0.1$ for stars having masses $\gtrsim 7 M_\odot$. Sagar et al. (1986) and Sagar \& Richtler (1991) also made similar conclusions.

High-mass stars have strong influence on their nearby surroundings and  can significantly affect the formation of low mass stars. Recently a relatively large number of low mass stars have been detected in a few OB 
associations, e.g. Upper Scorpios, the $\sigma$ and $\lambda$ Ori regions (Preibisch \& Zinnecker 1999, Dolan \& Mathieu 2002). Since this realization, surveys have demonstrated that the IMFs must be essentially the same in all star forming regions (e.g. Preibisch \& Zinneker 1999, Hillenbrand 1997, Massey et al. 1995a). The apparent difference is due mainly to the inherent low percentage of high mass stars and the incomplete survey of low mass stars in high-mass star forming regions. Advancement in detectors along with various surveys such as the 2MASS, DENIS and ISO have permitted detailed studies of low-mass stellar population in regions of high mass star formation.

The very young open cluster NGC 1893 (Massey et al. 1995a) is considered to be the center of the Aur OB2 association. NGC 1893 can be recognized as an extended region of loosely grouped early-type stars, associated with the H II region IC 410 with two pennant nebulae, Sim 129 and Sim 130 (Gaze \& Shajn 1952) and obscured by several conspicuous dust clouds. NGC 1893 contains at least five O-type stars, two of which, HD 242908 and LS V $+33^\circ16$ (S3R2N15) are main-sequence O5 stars, and therefore younger than $\sim 3$ Myr (Marco \& Negueruela 2002).

$UBV$ photometry of NGC 1893 has been presented by Hoag et al. (1961), Cuffey (1973), Moffat \& Vogt (1974) 
and Massey et al. (1995a). Tapia et al. (1991) performed near-infrared and Str\"{o}mgren photometry for 47 stars in the field of the cluster. They estimated the age of the cluster as 4 Myr and derived a distance modulus 
$(m-M)_0 = 13.18\pm0.11$ mag (4.3 kpc) and an extinction $A_V$ = 1.68 mag towards the cluster region. 
Str\"{o}mgren photometry for 50 stars in the field of NGC 1893 has also been 
reported by Fitzsimmons (1993), who confirmed the distance and age found by Tapia et al. (1991). Vallenari et al. (1999) performed near-infrared photometry of the 
cluster and came to the conclusion that there could be many pre-main sequence
candidates in NGC 1893, although their method did not allow clear discrimination from field interlopers. Cuffey (1973) derived $E(B-V)=0.4$ mag and distance = 3.6 kpc for the cluster.
Marco et al. (2001) derived an average distance modulus $V_0 - M_V = 13.9\pm0.2$ mag
(6 kpc) and $A_V = 1.42\pm0.13$ mag for the cluster, and also identified five emission-line stars as likely 
PMS members of NGC 1893. Recently they identified 18 emission-line PMS stars in NGC 1893 region (Negueruela et al. 2007).

The detection of PMS objects in NGC 1893 is interesting because it is one of the youngest known open cluster and has a moderately large population of O-type stars, representing thus a good laboratory for the study of massive star formation and the impact of massive stars on the formation of lower-mass stars (Marco \& Negueruela 2002). In this paper, we present a multi-wavelength study of the NGC 1893 to make a comprehensive exploration of the effects of massive stars on low mass star formation.
Deep optical $UBVRI$ and narrow band $H\alpha$ photometric data, slitless spectroscopy along with archival data from the surveys such as 2MASS, MSX, IRAS and NVSS are used to understand the global scenario of star formation in the NGC 1893 region.

\section{OBSERVATIONS AND DATA REDUCTION}

%-------------Fig 1----------------------------------------------------------------------------
\begin{figure*}
\centering
\includegraphics[height=15cm,width=16cm]{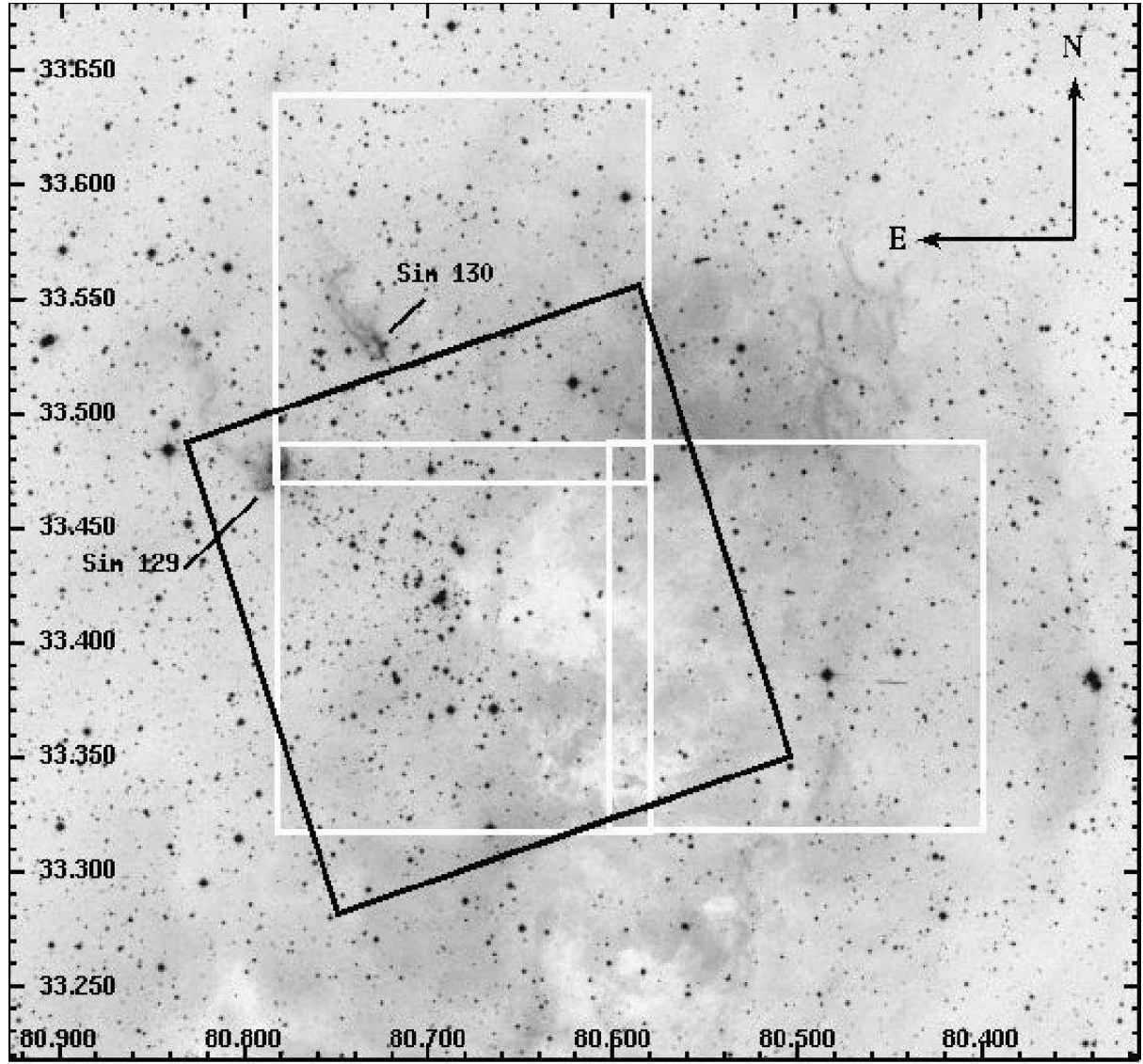}
\caption{DSS image of the region around NGC 1893 cluster. The black square box represents the observed region 
of cluster in optical bands. The areas marked by the white boxes show the region of slitless 
spectroscopic observations. The X and Y axes are RA and DEC coordinates in degrees in J2000 epoch.}
\end{figure*}
%----------------------------------------------------------------------------------------------

\subsection{Optical data}

The CCD $UBV{(RI)}_c$ and $H\alpha$+continuum photometric data were acquired on 06 and 18 January 2005 respectively using the $2048\times 2048$ pixel$^2$ CCD camera mounted on the f/13 Cassegrain focus of the 104-cm Sampurnanand telescope of Aryabhatta Research Institute of Observational Sciences (ARIES), Nainital. In this set up, each pixel of the CCD corresponds to $0.37$ arcsec and the entire chip covers a field of $\sim 13\times13$ arcmin$^2$ on the sky. To improve the signal to noise ratio, the observations were carried out in the binning mode of $2\times2$ pixel. The 
FWHMs of the star images were $\sim2$ arcsec. The read-out noise and gain of the CCD are 5.3 $e^-$ and 10 $e^-$/ADU respectively. The broad-band $UBV{(RI)}_c$ observations were standardized by observing stars in the SA98 field (Landolt 1992) on 07 January 2005. The observed region of NGC 1893 is shown in Fig. 1 as the black square box and the log of the observations is given in Table 1. A blank field of $\sim 13\times13$ arcmin$^2$ located at a distance of about $1^\circ$ away towards east of the cluster was also observed to estimate the contamination due to foreground/background field stars.

\begin{table}
\centering
\begin{minipage}{140mm}
\caption{Log of observation.}
\begin{tabular}{@{}rr@{}}
\hline
Date of observation/Filter& Exp. (sec)$\times$ No. of frames\\
\hline
&Sampurnanand telescope, ARIES\\
06 January 2005\\
$U$   &  $1200\times3,900\times1,300\times1,120\times3$\\
$B$   &  $600\times4,30\times3$\\
$V$   &  $600\times4,30\times4$\\
$R_c$   &  $300\times4,30\times1,10\times4$\\
$I_c$   &  $300\times4,30\times1,10\times4$\\
\\
18 January 2005\\
$H\alpha$ &$1200\times2,300\times3,60\times3$\\
$Continuum$&$1200\times2,300\times3,60\times3$\\
\\
&Himalayan Chandra Telescope, IIA\\
24 January 2006\\
Slitless spectra & $300\times9$ \\
Direct Frames  & $60\times3$\\
\hline
\end{tabular}
\end{minipage}
\end{table}

\begin{figure}
\centering
\includegraphics[height=7cm,width=8.5cm]{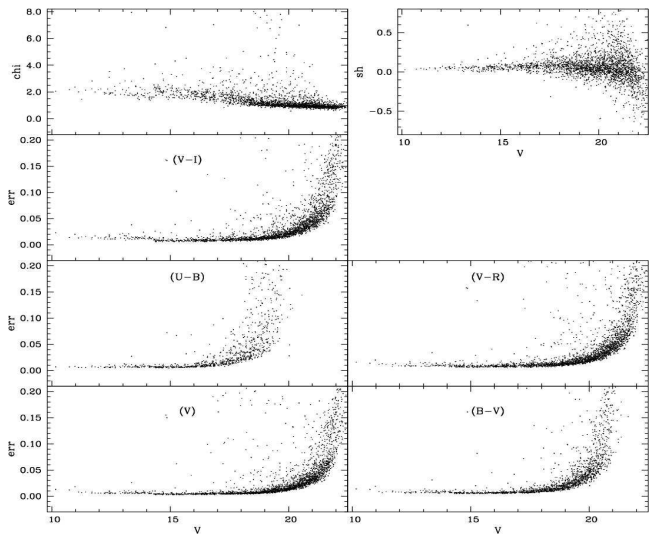}
\caption{The daophot errors in magnitude \& colours, image parameter $\chi$ and sharpness as a function
of $V$ magnitude.  }
\end{figure}

The CCD data frames were reduced using computing facilities available at
ARIES, Nainital.
Initial processing of the data frames were done
using the IRAF\footnote{IRAF is distributed by National Optical Astronomy
Observatories, USA} and ESO-MIDAS\footnote{ ESO-MIDAS is developed and 
maintained by the  European Southern Observatory.} data reduction packages. Photometry  of
cleaned frames was carried out using DAOPHOT-II software (Stetson 1987).
The PSF was obtained for each frame using several uncontaminated
stars. Magnitudes obtained from different frames
were averaged. When brighter stars were saturated on deep exposure frames, their
magnitudes have been taken from short exposure frames.
We used DAOGROW program for construction of an aperture growth curve required for
determining the difference between aperture and profile fitting magnitude.
Calibration of the instrumental magnitude to the standard system was
done by using procedures outlined by Stetson (1992).

For translating the instrumental magnitude to the standard magnitude, the calibration equations
derived using least-squares linear regression are as follows:\\

{\it  \small
\noindent
 $u= U + (7.004\pm0.004) -(0.005\pm0.006)(U-B) + (0.431\pm0.005)X$,
                                                                
\noindent                                                  
 $b= B + (4.742\pm0.005) -(0.035\pm0.004)(B-V) + (0.219\pm0.004)X$,
                                                                
\noindent                                                       
 $v= V + (4.298\pm0.002) -(0.038\pm0.002)(V-I) + (0.128\pm0.002)X$,
                                                                
\noindent                                                       
 $r= R + (4.202\pm0.004) -(0.046\pm0.007)(V-R) + (0.078\pm0.003)X$,
                                                                
\noindent                                                       
 $i= I + (4.701\pm0.004) -(0.059\pm0.003)(V-I) + (0.044\pm0.003)X$ \\

}

where $U,B,V,R$ and $I$ are the standard magnitudes and $u,b,v,r$ and $i$ are the
instrumental aperture magnitudes normalized for 1 second of exposure time and $X$
is the airmass. We have ignored the second-order colour correction terms as they are generally
small in comparison to other errors present in the photometric data reduction.
The typical DAOPHOT errors in magnitude and colour as a function of $V$ magnitude,
are shown in Fig. 2 and statistical results are given in Table 2. 
The parameter $\chi$ and sharpness are also shown in Fig. 2 as a function of $V$ magnitude. It can be seen that the errors become large ($\ge$0.1 mag) for stars
fainter than $V\simeq22$ mag, so the measurements beyond this magnitude are not reliable.
The standard deviations in the standardization residual, $\Delta$, between standard and transformed V magnitude and $(U-B),(B-V),(V-R)$ and $(V-I)$ colours of standard stars are 0.006, 0.025, 0.015, 0.010, 0.015 mag respectively.

\begin{table}
\centering
\begin{minipage}{140mm}
\caption{ Average photometric errors $\sigma$ as a function of brightness.}
\begin{tabular}{@{}rrrrrr@{}}
\hline
$V$ Magnitude &$\sigma_V$&$\sigma_{(U-B)}$&$\sigma_{(B-V)}$&$\sigma_{(V-R)}$&$\sigma_{(V-I)}$\\
Range           &&&&&\\
\hline
%$<10$           &    0.12  &      0.03      &    0.13      &    0.15      &    0.13\\
10 - 11         &    0.01  &      0.01      &    0.01      &    0.02      &    0.02\\
11 - 12         &    0.01  &      0.01      &    0.01      &    0.01      &    0.01\\
12 - 13         &    0.01  &      0.01      &    0.01      &    0.01      &    0.01\\
13 - 14         &    0.01  &      0.01      &    0.01      &    0.01      &    0.01\\
14 - 15         &    0.01  &      0.01      &    0.01      &    0.01      &    0.01\\
15 - 16         &    0.01  &      0.01      &    0.01      &    0.01      &    0.01\\
16 - 17         &    0.01  &      0.01      &    0.01      &    0.01      &    0.01\\
17 - 18         &    0.01  &      0.03      &    0.01      &    0.01      &    0.01\\
18 - 19         &    0.01  &      0.04      &    0.02      &    0.02      &    0.01\\
19 - 20         &    0.02  &      0.07      &    0.04      &    0.02      &    0.02\\
20 - 21         &    0.03  &                &    0.06      &    0.04      &    0.04\\
21 - 22         &    0.05  &                &    0.08      &    0.06      &    0.06\\
\hline
\end{tabular}
\end{minipage}
\end{table}

\subsection{Grism slitless spectroscopy}

\begin{figure}
\centering
\includegraphics[height=7cm,width=8.5cm]{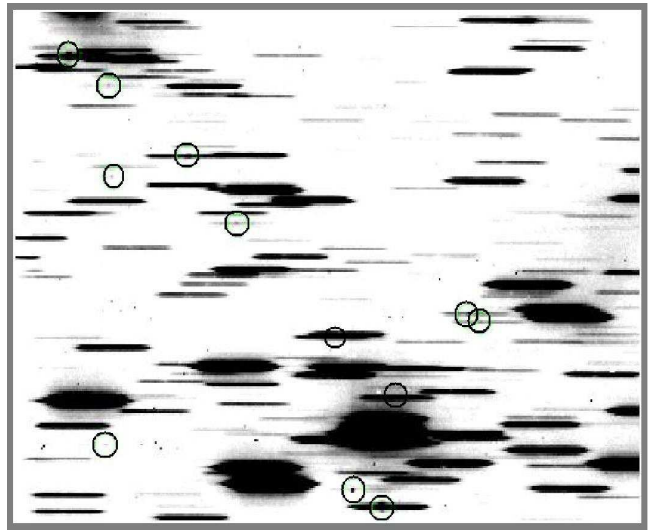}
\caption{One of the slitless spectra of the cluster region. The $H\alpha$ emission line sources have
been marked by circles. }
\end{figure}

Three regions of the cluster (marked with white square boxes in Fig. 1) 
were also observed in the slitless mode with a grism as the dispersing element using the Himalayan Faint Object Spectrograph Camera (HFOSC) instrument on 24 January 2006. This yields an image where the
stars are replaced by their spectra. A combination of a `wide H$\alpha$' interference filter 
(6300 - 6740 \AA) and Grism 5 (5200 - 10300 \AA) of HFOSC was used without any slit. The
resolution of grism is 870. The central $2K\times2K$ pixels of the $2K\times4K$ CCD
were used for imaging. The pixel size is 15 micron with an image scale of 0.297 arcsec/pixel.
For each region we secured three spectroscopic frames of 5 min exposure each with the grism in, and one direct frame of 1 min exposure with the grism out. 
Emission line stars with enhancement over the continuum at H$\alpha$ wavelength are visually identified. Fig. 3 shows the slitless spectra of stars in one of the cluster regions. Positions of the H$\alpha$ emission stars are given in Table 3.
Recently Negueruela et al. (2007) have reported detection of 18 $H\alpha$ emission stars, of which 12 lies in the region surveyed by us. However only 9 stars are common in both the surveys.

\begin{figure*}
\centering
\hbox{
\includegraphics[height=5cm,width=6cm]{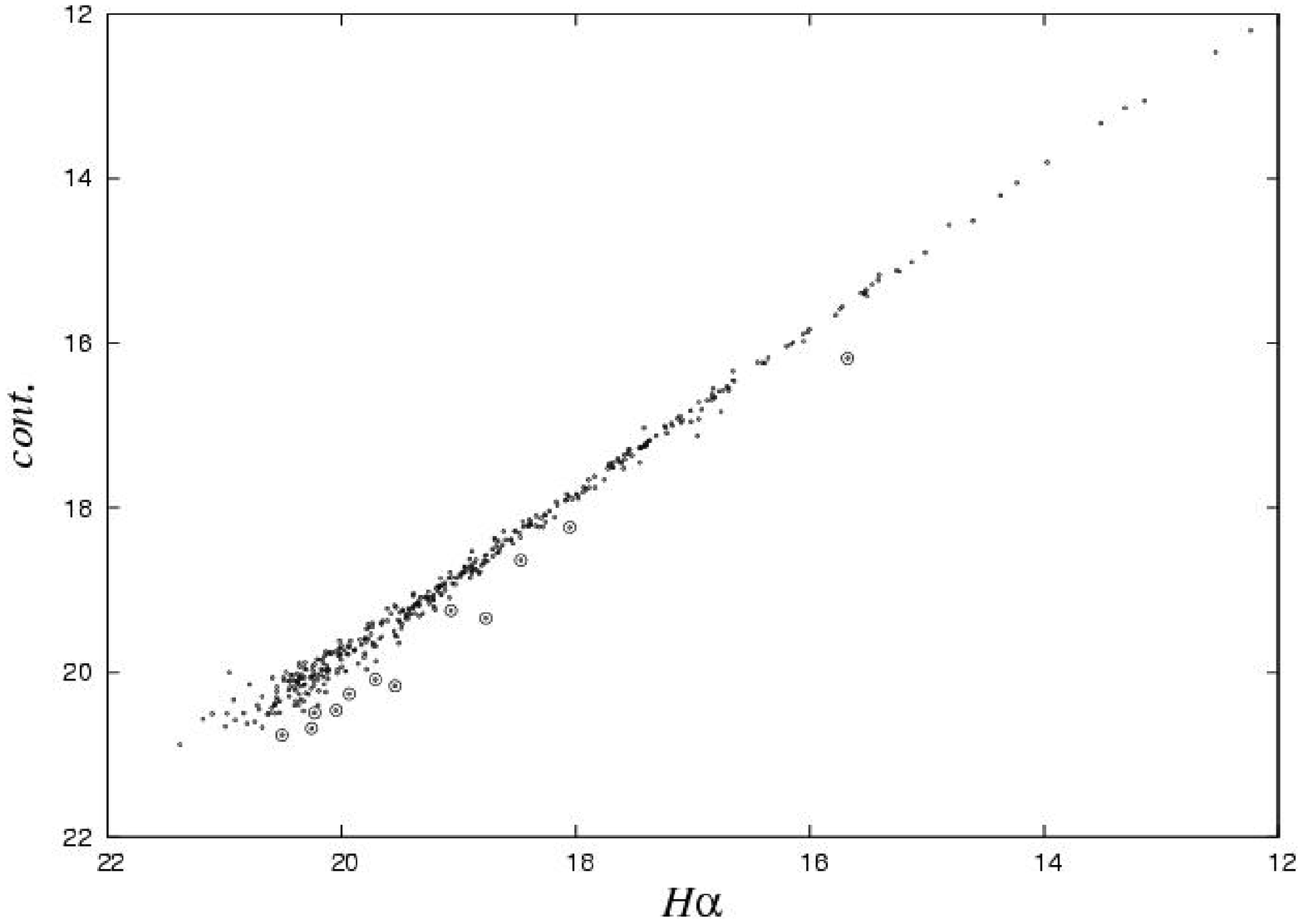}
\includegraphics[height=5cm,width=6cm]{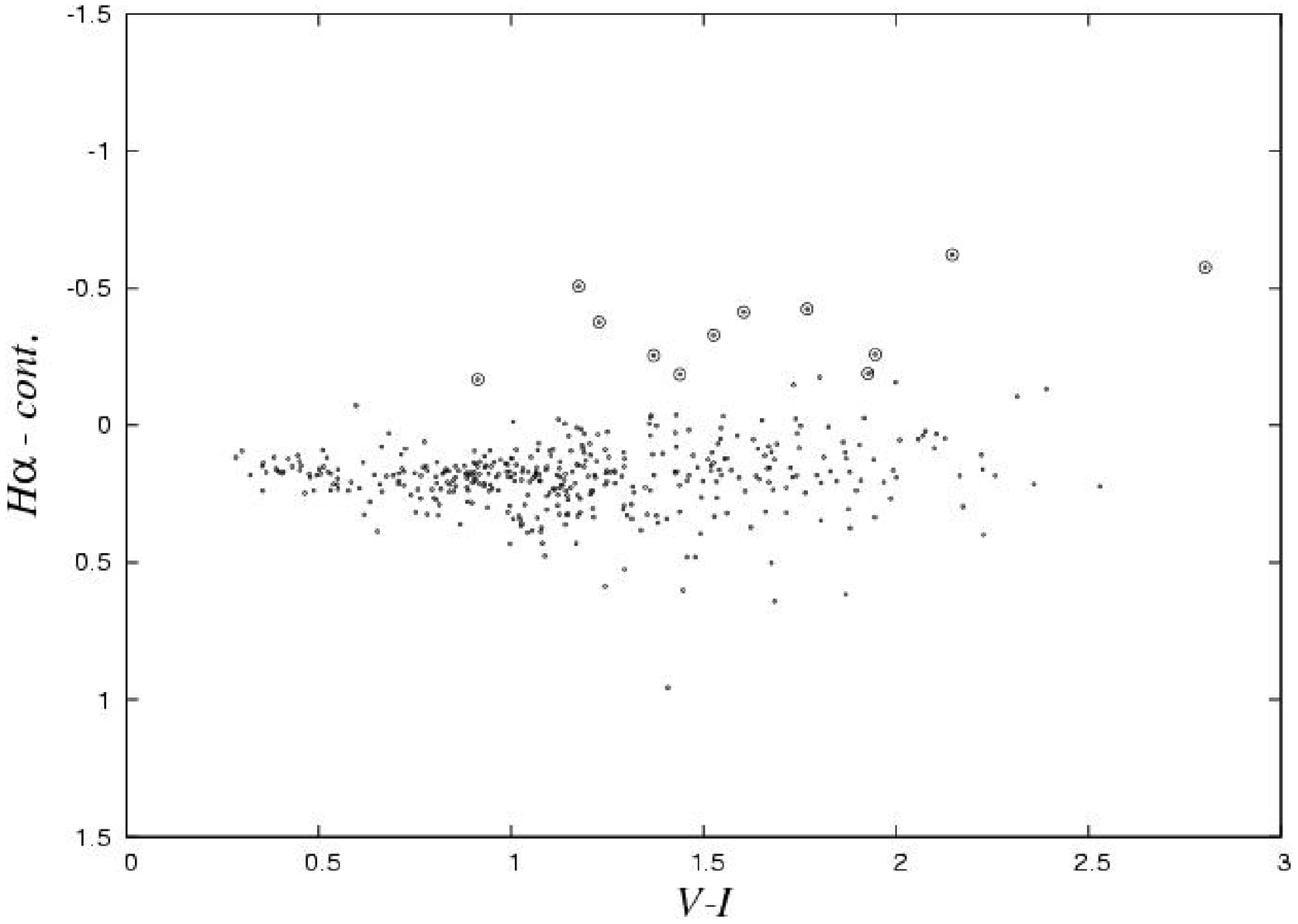}
\includegraphics[height=5cm,width=6cm]{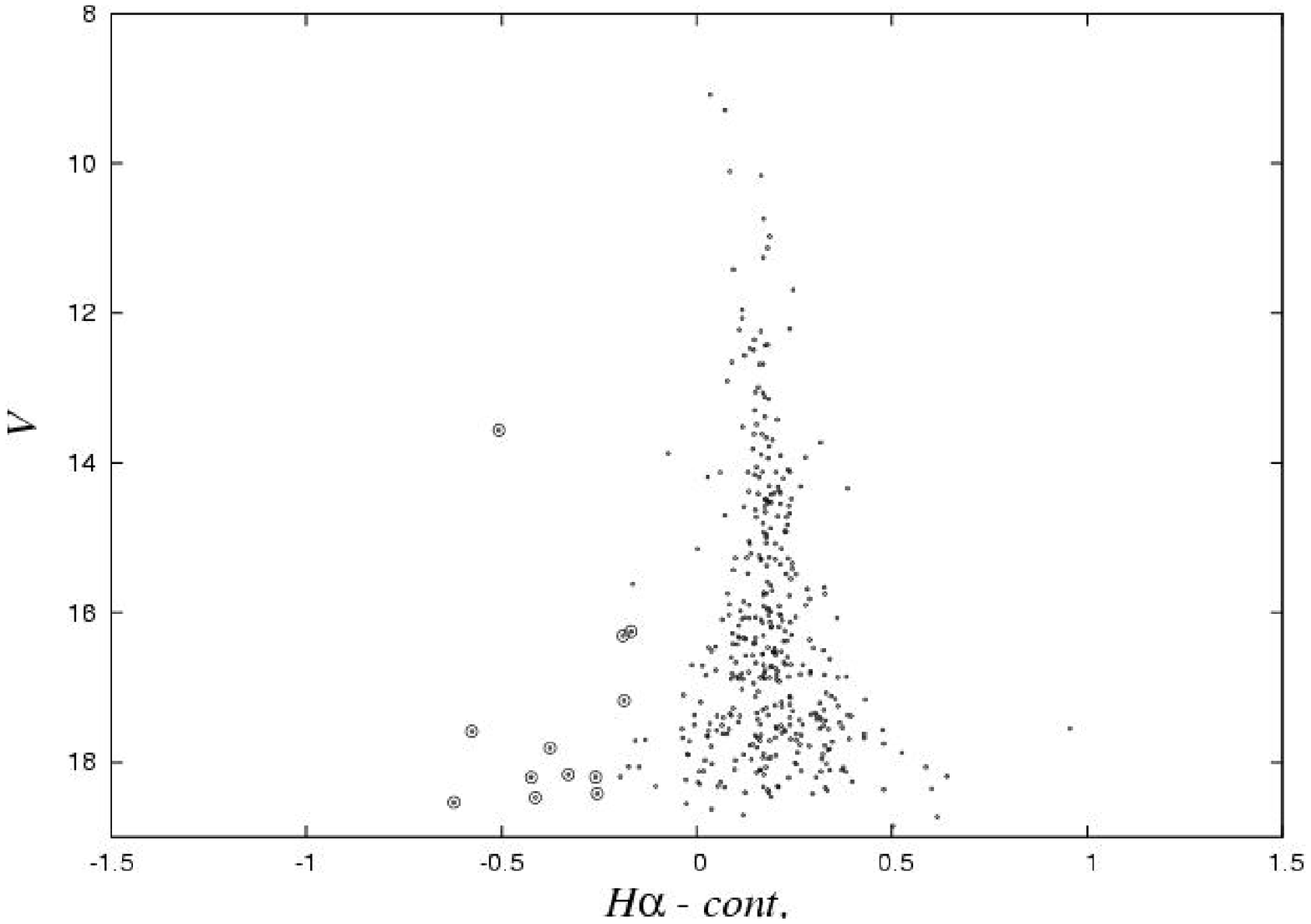}
}
\caption{Left panel: $H\alpha$ vs $cont.$ diagram, middle panel: $(H\alpha - cont.)$, $(V-I)$ diagram and right panel: $V,(H\alpha - cont.)$. Circle represents probable $H\alpha$ emission stars. }
\end{figure*}

\subsection{H$\alpha$ photometry}

The excess $H\alpha$ emission stars may also be detected by imaging the cluster region in $H\alpha$ line and a nearby continuum (see e.g. Sung et al. 2000). We calculated the difference $\Delta$ =($H\alpha_{instrumental} - Continuum_{instrumental}$) and designated a star having excess emission if $\Delta$ is greater than three sigma from the mean value of the distribution in the $H\alpha / Cont.$ diagram. In Fig. 4, we have plotted $H\alpha/Cont.$, $(H\alpha-Cont.)/(V-I)$ and $V/(H\alpha-Cont.)$ diagrams. Probable $H\alpha$ emission stars are shown by open circles. Information about photometrically detected $H\alpha$ emission stars is given in Table 3. Only 3 probable photometrically detected  $H\alpha$ emission stars are common with those detected by the slitless spectroscopy. Here it is worthwhile to point out that some late-type dwarfs show strong chromospheric $H\alpha$ in emission. Huang et al. (2006) have pointed out that, even if a late-type star does not show $H\alpha$ emission, such a star generally has a series of strong metal oxide absorption lines, such as TiO in its spectrum. This may lead to an underestimation of their continuum, hence making the star as a $H\alpha$ emitter.

\begin{table}
\centering
\begin{minipage}{140mm}
\caption{Position of all the $H\alpha$ emission stars.}
\begin{tabular}{@{}rrrrrrrr@{}}
\hline
    \multicolumn{3}{c}{slitless spectroscopy}  & \multicolumn{3}{c}{ narrow band photometry}\\
ID &   RA(2000)    &  DEC(2000)     &ID &   RA(2000)     &  DEC(2000)  \\
   &  (h:m:s)    & (d:m:s)      &   & (h:m:s)      & (d:m:s)\\
\hline

 A&05:22:43.02&  +33:25:05.4&    a& 05:22:19.86&  +33:22:14.2\\
 B&05:22:43.78&  +33:25:25.8&    b& 05:22:19.90&  +33:21:32.5\\
 C&05:22:45.78&  +33:28:16.2&    c& 05:22:22.41&  +33:20:18.0\\
 D&05:22:46.08&  +33:24:57.8&    d& 05:22:27.64&  +33:20:13.5\\
 E&05:22:49.57&  +33:30:01.5&    e& 05:22:30.67&  +33:19:27.1\\
 F&05:22:51.04&  +33:25:47.1&    f& 05:22:31.14&  +33:22:58.9\\
 G&05:22:51.90&  +33:23:59.3&    g& 05:22:31.92&  +33:29:05.7\\
 H&05:22:52.23&  +33:29:58.0&h$^*$& 05:22:43.02&  +33:25:05.4\\
 I&05:22:52.30&  +33:24:07.4&i$^*$& 05:22:43.78&  +33:25:25.8\\
 J&05:22:56.42&  +33:26:47.7&    j& 05:22:46.84&  +33:29:28.1\\
 K&05:22:58.10&  +33:30:41.0&k$^*$& 05:22:49.57&  +33:30:01.5\\
 L&05:22:58.85&  +33:28:34.6&    l& 05:22:51.23&  +33:20:35.5\\
 M&05:22:59.57&  +33:27:32.5&     &            &\\
 N&05:23:00.07&  +33:30:39.0&     &            &\\
 O&05:23:02.87&  +33:28:17.8&     &            &\\
 P&05:23:04.26&  +33:28:46.4&     &            &\\

\hline
\end{tabular}
\end{minipage}
$^*$: also detected  in slitless spectroscopy\\
\end{table}

\section{Other available data sets}

Near-IR (NIR) ($JHK_s$) data for point sources around the cluster region have been obtained from the 
Two Micron All Sky Survey (2MASS) Point source Catalog (PSC). 
The 2MASS data base provides
photometry in the near infrared $J(1.25\mu$m), $H(1.65 \mu$m) and $K_s (2.17 \mu$m)
bands to a limiting magnitude of 15.8, 15.1 and 14.3 respectively, with a signal to noise
ratio (S/N) greater than 10. We retain only those sources for which the error in each band is less than 0.1 mag to ensure good photometric accuracy.

The Midcourse Space experiment (MSX) surveyed the Galactic plane in four mid-infrared bands  - A ($8.28~ \mu$m), C ($12.13~ \mu$m), D ($14.65~ \mu$m) and E ($21.34~ \mu$m) at a spatial resolution of $\sim18^{\prime \prime}$ (Price et al. 2001). Two of these bands (A and C) with ${\it \lambda(\Delta \lambda)}$ corresponding to 8.28(3.36) and 12.13(1.71) include several Unidentified Infrared emission Bands (UIBs) at 6.2, 7.7, 8.7 11.3, and 12.7 $\mu$m. MSX images in these four bands around the cluster region were used to study the emission from the UIBs and to estimate the spatial distribution
of temperature and optical depth of  warm interstellar dust.

The data from the IRAS survey around the cluster region in the four bands
(12, 25, 60, 100 $\mu$m) were HIRES processed (Aumann et al. 1990) to obtain 
high angular resolution maps. These maps were used to determine the spatial 
distribution of dust colour temperature and optical depth. 
Six IRAS point sources have also been identified in the cluster region and their details are given in Table 4.

\begin{table*}
\centering
\begin{minipage}{140mm}
\caption{List of IRAS point sources.}
\begin{tabular}{@{}rrrrrrrr@{}}
\hline

     IRAS PSC & RA(2000)&DEC(2000)& $F_{12}$   &    $F_{25}$&      $F_{60}$&     $F_{100}$\\
            &(degrees)&(degrees)&          (Jy)&          (Jy)&          (Jy)&          (Jy)\\
\hline
  05186+3326& 80.4944 & 33.4943 &    1.47 &    2.12 &    22.95 &    146.60\\
  05189+3327& 80.5674 & 33.5059 &    1.31 &    2.06 &    46.35 &    120.50\\
  05194+3322& 80.6848 & 33.4240 &    1.16 &    1.38 &     2.29 &     29.48\\
  05196+3329& 80.7390 & 33.5326 &    1.08 &    2.39 &    28.30 &    163.40\\
  05198+3325& 80.7844 & 33.4771 &    8.33 &   26.16 &   145.70 &    163.40\\
  05200+3329& 80.8315 & 33.5299 &    1.39 &    1.91 &   145.70 &    163.40\\
\hline
\end{tabular}
\end{minipage}
\end{table*}

\section{ COMPARISON WITH PREVIOUS STUDIES}

We have carried out a comparison of the present photometric data with those
available in the literature. The difference $\Delta$ (literature -
present data) as a function of $V$ magnitude is shown in Fig. 5 and
the statistical results are given in Table 5. The comparison
indicates that the present photometry is in agreement with the CCD photometry by Massey et al. (1995a) and photoelectric photometry by Hoag et al. (1961). The $(B-V)$ and $(U-B)$ colours by Cuffey (1973) are in agreement with the present photometry whereas the $\Delta V$ shows some variation. 

\begin{table*}
\centering
\begin{minipage}{140mm}
\caption{\label{Table:4}Comparison of the present photometry with the available photometry in the literature. The difference $\Delta$ (literature-present data) is in magnitude. Mean and $\sigma$ are based on N stars in a V magnitude bin.}
\begin{tabular}{@{}lrrrrrr@{}}
\hline
$V$ range&$\Delta(V)$ && $\Delta(B-V)$&&$\Delta(U-B)$&\\
&($Mean\pm \sigma$)&(N) &($Mean\pm \sigma$)&(N)&( $Mean\pm \sigma$)&(N)\\
\hline
Massey P. et al. (1995a, ccd)\\
 10-11&$  0.033\pm0.010$& 2   &$ -0.012\pm0.016$&2  &$  0.006\pm0.022$&2\\
 11-12&$  0.042\pm0.020$& 6   &$ -0.010\pm0.010$&6  &$ -0.045\pm0.022$&6\\
 12-13&$  0.036\pm0.027$& 16  &$ -0.003\pm0.021$&16 &$ -0.022\pm0.033$&16\\
 13-14&$  0.035\pm0.040$& 16  &$  0.001\pm0.022$&16 &$ -0.012\pm0.051$&16\\
 14-15&$  0.020\pm0.075$&43   &$  0.029\pm0.077$&43 &$ -0.010\pm0.030$&43\\
 15-16&$  0.003\pm0.092$&55   &$  0.024\pm0.070$&54 &$  0.015\pm0.038$&53\\
 16-17&$ -0.003\pm0.074$&109  &$  0.032\pm0.087$&109&$  0.015\pm0.103$&106\\
 17-18&$  0.004\pm0.010$&73   &$  0.015\pm0.136$&73 &$  0.023\pm0.189$&71\\
\hline
Cuffey J. (1973, pe)\\
 10-11&$  0.043\pm0.004$& 2   &$  0.002\pm0.020$&2  &$  0.017\pm0.021$&2\\
 11-12&$  0.120\pm0.024$& 3   &$ -0.023\pm0.034$&3  &$ -0.005\pm0.006$&3\\
 12-13&$  0.095\pm0.052$& 10  &$ -0.036\pm0.036$&10 &$ -0.005\pm0.036$&10\\
 13-14&$ -0.064\pm0.244$& 5   &$ -0.004\pm0.151$&5  &$ -0.033\pm0.038$&5\\
 14-15&$ -0.066\pm0.014$&3    &$  0.049\pm0.062$&3  &$ -0.042\pm0.043$&3\\
 15-16&$ -0.035\pm0.023$&2    &$  0.063\pm0.019$&2  &$  0.049\pm0.066$&2\\
\hline
Hoag et al. (1961, pe)\\
 10-11&$  0.013\pm0.010$& 2   &$  0.007\pm0.016$&2  &$ -0.008\pm0.001$&2\\
 11-12&$  0.047\pm  -   $& 1   &$  0.006\pm  -   $&1  &$ -0.037\pm  -   $&1 \\
 12-13&$  0.033\pm0.042$& 7   &$  0.004\pm0.032$&7  &$ -0.036\pm0.044$&7\\
 13-14&$  0.020\pm0.048$& 6   &$  0.030\pm0.022$&6  &$ -0.018\pm0.070$&6\\
 14-15&$ -0.017\pm0.047$&3    &$  0.002\pm0.011$&3  &$  0.003\pm0.032$&3\\
 
\hline

\end{tabular}
\end{minipage}
\end{table*}

\section{COMPLETENESS OF THE DATA}

To study the luminosity function  (LF)/ mass function (MF) it is very important to make necessary corrections in data sample to take into account the incompleteness that may occur for various reasons (e.g. crowding of the stars). We used the ADDSTAR routine of DAOPHOT II to determine the completeness factor (CF). The 
procedures have been outlined in detail in our earlier work (Sagar \& Richtler 1991, Pandey et al. 2001, 2005). In the case of optical CCD photometry the incompleteness of the data increases with increasing magnitude as expected. The completeness factor (CF) as a function of $V$ magnitude is given in Table 6. 
In the case of 2MASS the data is found to be almost complete up to 14 mag in the $K_{\rm s}$ band.

\begin{figure}
\centering
\includegraphics[height=4cm,width=8cm]{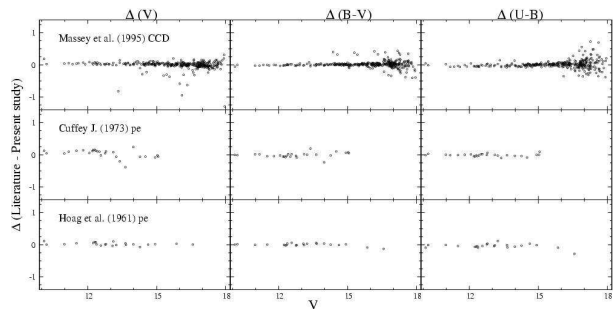}
\caption{Comparison of present data with those given in literature. The difference
($\Delta$ = present study - literature) is plotted as a function of $V$ magnitude.}
\end{figure}

\begin{table}
\centering
\begin{minipage}{80mm}
\caption{Completeness Factor (CF) of photometric data in the cluster and field regions.}
\begin{tabular}{@{}rrrr@{}}
\hline
V range &\multicolumn{2}{c}{ NGC 1893}  & Field region \\

(mag)& $r\le2^\prime$ & $2^\prime<r\le6^\prime$&\\
\hline

10 - 11   &    1.00      &    1.00&       1.00\\
11 - 12   &    1.00      &    1.00&       1.00\\
12 - 13   &    1.00      &    1.00&       1.00\\
13 - 14   &    1.00      &    1.00&       1.00\\
14 - 15   &    1.00      &    1.00&       1.00\\
15 - 16   &    1.00      &    1.00&       1.00\\
16 - 17   &    1.00      &    1.00&       1.00\\
17 - 18   &    1.00      &    1.00&       1.00\\
18 - 19   &    0.95      &    1.00&       1.00\\
19 - 20   &    0.96      &    0.93&       0.96\\
20 - 21   &    0.68      &    0.65&       0.68\\
\hline
\end{tabular}
\end{minipage}
\end{table}

\section{STRUCTURE OF THE CLUSTER}

\begin{figure*}
\centering
\hbox{
\includegraphics[height=8cm,width=7cm]{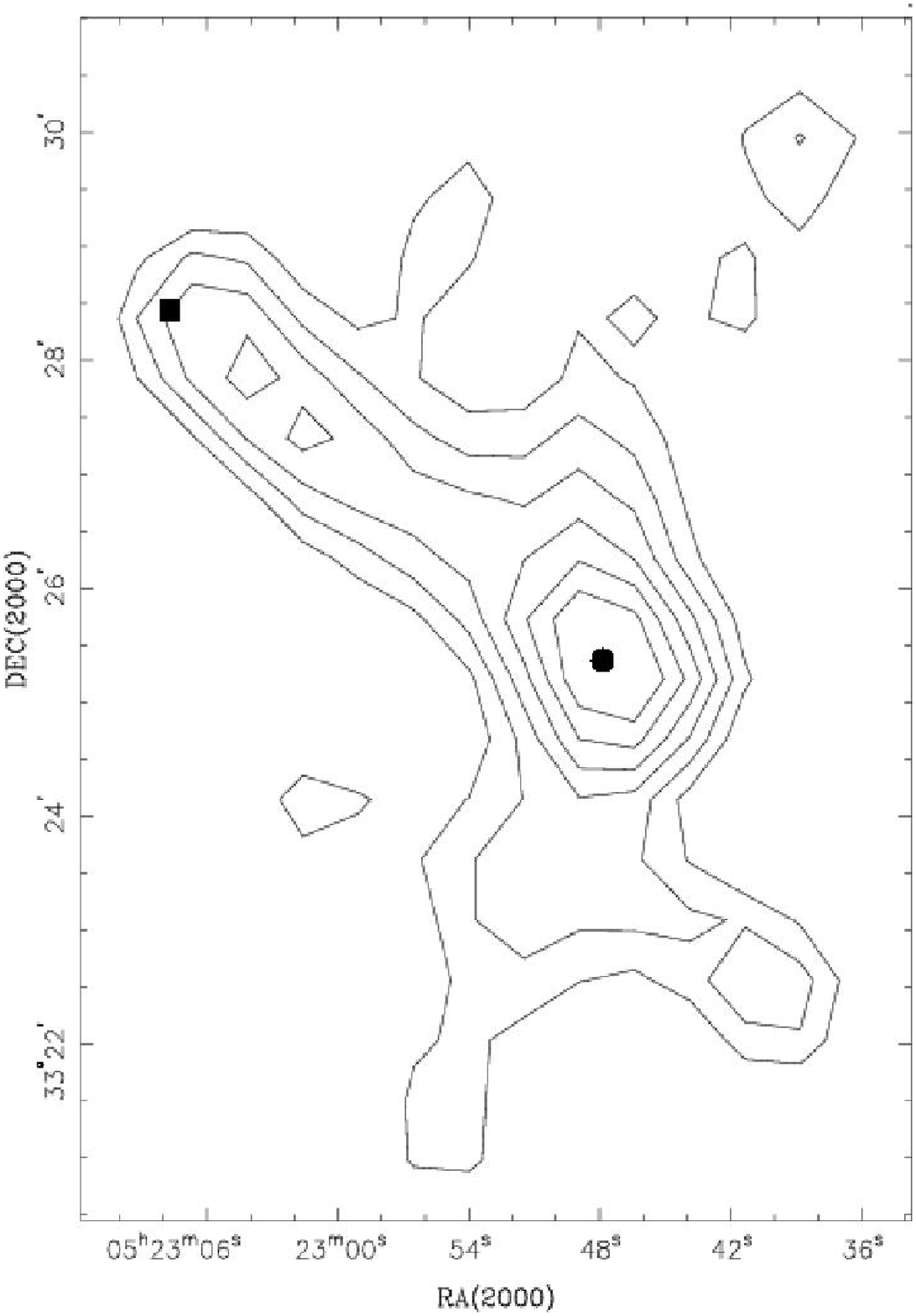}
\hspace{1cm}
\includegraphics[height=8cm,width=7cm]{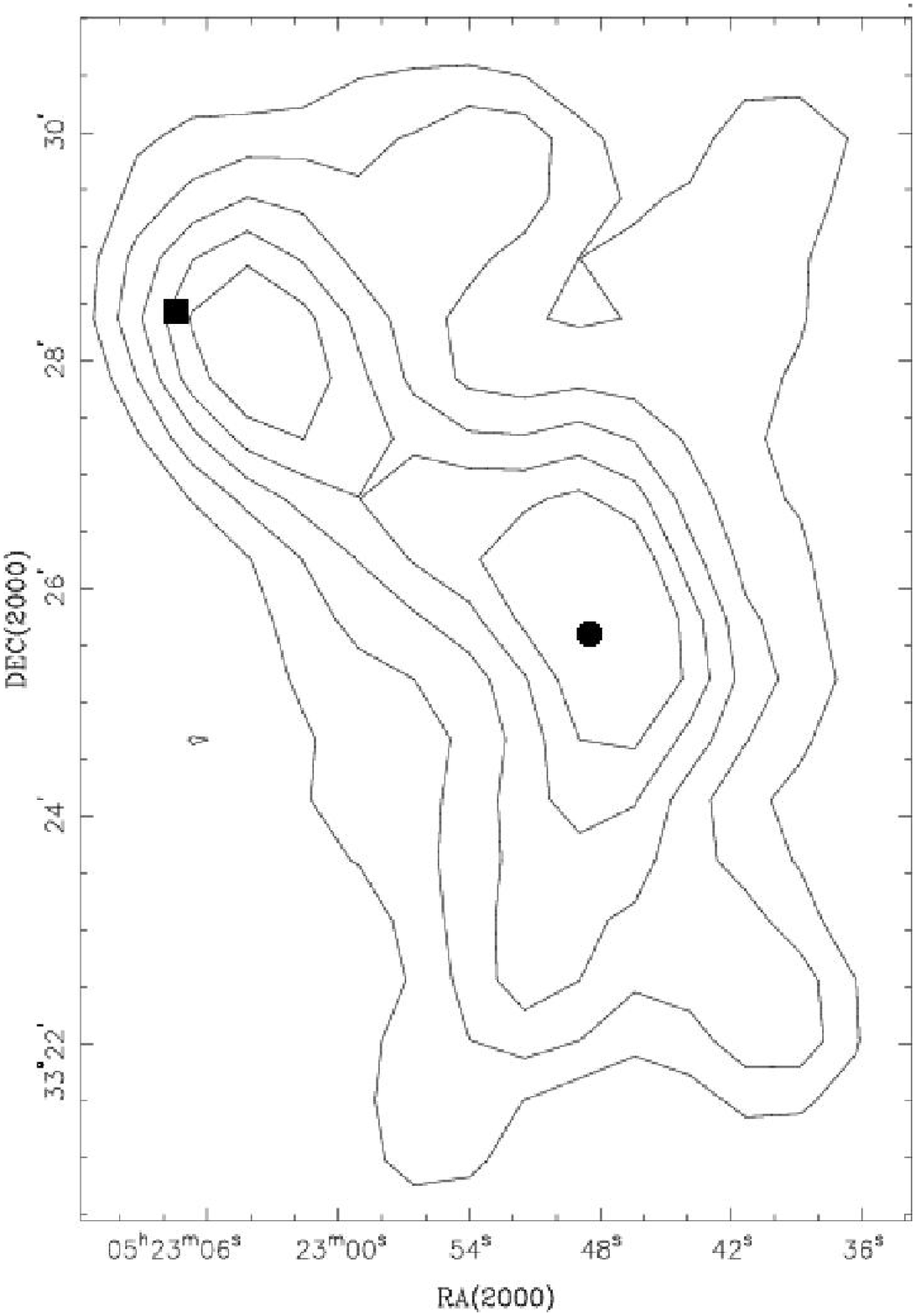}
}
\caption{Isodensity contours for (a) stars having $V\le18$ mag; (b) 2MASS stars in $K_{\rm s}$ band.
The contours are plotted above 3 sigma levels. 
For optical data, the contours have step size of 1 $star/pc^2$ with the lowest contour representing 
5 $stars/pc^2$. The maximum stellar density is ~12 $stars/pc^2$.
For 2MASS data, the contours have step size of 2 $stars/pc^2$ with lowest contour representing 
6 $stars/pc^2$. The maximum stellar density is ~18 $stars/pc^2$. 
The filled square and circle represent the locations of nebula Sim 129 and cluster center.}
\end{figure*}

\subsection{Isodensity contours}

The initial stellar distribution in star clusters may be governed by the structure of parental molecular cloud and also how star formation proceeds in the cloud (Chen et al. 2004, Sharma et al. 2006). Later evolution of the cluster may be governed by internal gravitational interaction  among member stars and external tidal forces due to the Galactic disk or giant molecular clouds. 

To study the morphology of the cluster, we have generated isodensity contours (Fig. 6) for stars from 
optical ($V\le18$) as well as from NIR 2MASS data.
The contours are plotted above the 3-sigma value of the background level as estimated from the control field. The filled square and circle marked in Fig. 6 represent the location of the nebula Sim 129 and cluster center (as estimated in next section) respectively. The isodensity contours indicate that the cluster has elongated morphology.

\subsection{Radial density profile}

To estimate the radial extent of the cluster we assume a spherical symmetry for the cluster. The center of the cluster is determined using the stellar density distribution of stars having $V \leq 20$ mag, in a  $\pm$ 100 pixels wide strip along both X and Y directions around an initially eye estimated center. The point of maximum density obtained by fitting a Gaussian curve, is considered as the center of the cluster. The (X, Y) pixel
coordinates of the cluster center are found to be  (585, 394)
which corresponds to $\alpha_{2000} = {05^h} {22^m} {49^s}$, $\delta_{2000} =
33^{\circ} 25^{'} 35^{''}$.

To determine the radial surface density we divided the cluster into
a number of concentric circles. Projected radial stellar density in each
concentric annulus was obtained by dividing the number of stars in each annulus
by its area and the same are plotted in Fig. 7 for various magnitude levels.
The error bars are derived assuming that the number of stars in a concentric  
annulus follow the Poisson statistics.  The horizontal dashed
line in the plot indicates the density of contaminating field stars, which is
obtained from the region $\sim 1^\circ$ away towards east of the cluster center. 

To check whether the density distribution shown in the left panel of Fig. 7
is affected by contamination due to field stars, we selected a sample of stars
near a well defined main sequence (MS) in the colour-magnitude diagram (CMD)
as mentioned by Pandey et al. (2001). The radial density distribution of the MS sample
is also shown in the right panel of Fig. 7. 
Both the samples in general show almost similar radial density profiles.
We have also used 2MASS data to obtain the radial density profile of the cluster and the same is shown in Fig. 7 (bottom panels), which also shows a similar radial density profile.

We have used these profiles to calculate the extent of cluster $`r_{cl}$' which is defined as the point at which the radial density becomes constant and merges with the field density. 
The radial distribution of stars shown in Fig. 7 indicates that in general the extent of 
NGC 1893 cluster is $\sim 6^{\prime}$ (5.7 pc).

To parametrize the radial density profiles we follow the approach by Kaluzny
 \& Udalski (1992). Because of the low S/N ratio in the star counts
of open clusters, it is not an easy task to constrain the tidal radius of cluster
using the empirical model of King (1962). We describe radial density $\rho (r)$ as

$$\rho (r) \propto {fo\over
\displaystyle{1+\left({r\over r_c}\right)^2}}\,;$$

 where $r_c$ is the core radius (the radius at which the surface density falls
to half of the central density $f_o$). We fit above function to the observed
radial density profile of stars. The best fit is obtained by $\chi^2$ minimization technique.
The fit was performed for the data within the radii of 10$^{\prime}$.

The core radius for various magnitude levels (14 $\le V \le$ 20) varies 
from $1.^{\prime}1\pm0.^{\prime}2$ ($\sim 1.0\pm0.2$ pc) to $2.^{\prime}6 \pm 0.^{\prime}7$ ($\sim 2.5\pm0.7$ pc), whereas for main sequence stars it varies from $1.^{\prime}0 \pm 0.^{\prime}2$  ($\sim 1.0\pm0.2$ pc) to $2.^{\prime}8 \pm 0.^{\prime}6$ ($\sim 2.6\pm0.6$ pc).  The 2MASS data gives a core radius as $1.^{\prime}6 \pm 0.^{\prime}6$ ($\sim 1.5\pm0.6$ pc)
for  both the samples. 
To study the colour-magnitude diagrams (CMDs), MF etc., in detail we divided the cluster into two subregions as  the inner region ($r< 2^{\prime} $, 1.9 pc), and the outer region (2$^{\prime} \leq r \leq 6^{\prime}$, 1.9 pc$\leq r \leq$ 5.7 pc).

\begin{figure}
\centering
\includegraphics[height=8cm,width=8cm,angle=-90]{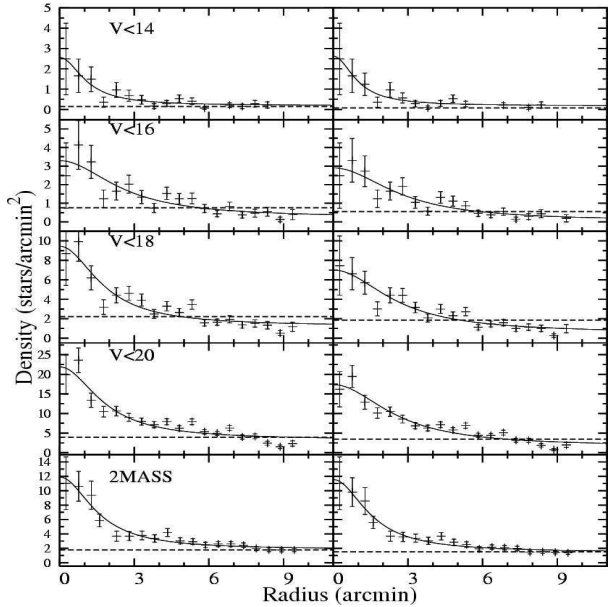}
\caption{The radial density profiles of the cluster NGC 1893 for different optical magnitude levels and 2MASS data 
(in K$_{\rm s}$ band) for all (left panels) and MS (right panels) stars. The solid curve shows a least-square fit of the King (1962) profile to the observed data points. 
The error bars represents $\pm \sqrt{N}$ errors. The dashed line indicate the density
of field stars. }
\end{figure}

\section {Interstellar Extinction}

\subsection{Reddening}

The extinction towards the cluster region is estimated using the $(U-B)/(B-V)$
two-color diagram (TCD) shown in Fig. 8, where the MS (Schmidt-Kaler, 1982) is shifted along the reddening vector having an adopted slope of $E(U-B)/E(B-V)$ = 0.72 to match the observations. The TCD shows a spread in the observed sequence indicating the presence of differential reddening which is an indication of the youth of the cluster. 
Fig. 8 yields a spread in reddening between $E(B-V)$ = 0.4 to 0.6 mag.
It  has already been noticed that the extinction towards the anticenter direction of the Galaxy is relatively low (cf. Pandey et al. 1989, Joshi 2005). Recent work by Joshi (2005) yields an average value of $E(B-V) \sim 0.45$ mag which is in agreement with the values ($E(B-V) = 0.4 - 0.6$ mag) obtained in the present work.

The reddening of individual stars ($V_{error}< 0.1$) having spectral type earlier than A0, has been derived using the $Q$ method (Johnson \& Morgan 1953). The distribution of reddening as a function of radial distance is shown in Fig. 9 which indicates a deficiency of reddening material in the central region of the cluster.

\begin{figure}
\centering
\includegraphics[height=8cm,width=9cm]{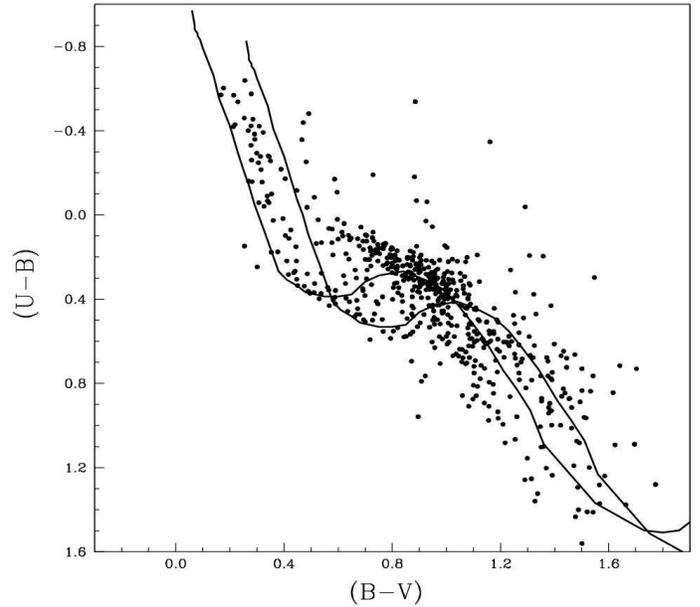}
\caption{The $(U-B)/(B-V)$ colour-colour diagram for the stars lying within the cluster region ($r<r_{cl}$). The continuous curve represents intrinsic MS by Schmidt Kaler (1982) shifted along the reddening vector of 0.72 for $E(B-V)_{min}$ = 0.4 and  $E(B-V)_{max}$ = 0.6. }
\end{figure}

\begin{figure}
\centering
%\hbox{
\includegraphics[height=8cm,width=5.2cm,angle=-90]{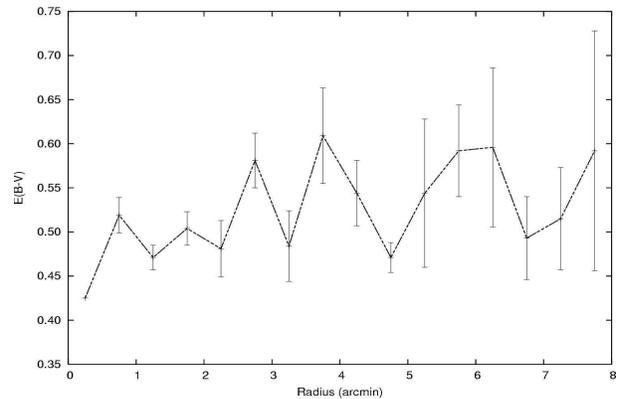}
\caption{Spatial variation of reddening in the cluster region. Error bars represent standard errors. }
%}
\end{figure}

\begin{figure}
\centering
\includegraphics[height=8cm,width=6cm,angle=-90]{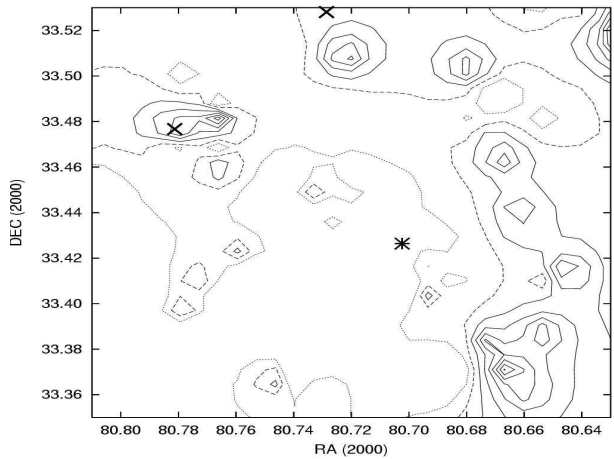}
\caption{Reddening distribution in the cluster region. The dotted, dashed and continuous contours are for $E(B-V)$=0.51, 0.54 and $\ge 0.57$ mag respectively. The cross and asterisk represent the locations of nebula Sim 129 and cluster center.}
\end{figure}

The contour maps of reddening in the cluster region are shown in Fig. 10, which indicates that reddening toward the west of the cluster is higher. Individual enhancements of reddening can be seen around some regions including the nebulae Sim 129 and Sim 130 (marked by crosses). The center of NGC 1893  (marked by asterisk) is less reddened and seems to be devoid of reddening material.

\subsection{Reddening law}

The interstellar extinction and the ratio of total-to-selective extinction $R = A_V / E(B-V)$ towards the cluster are important quantities that must be accurately known to determine the distance of the cluster photometrically.
To study the nature of the extinction law in the cluster region, we used TCDs as described by Pandey et al. (2003). The TCDs of the form of ($V-\lambda$) vs. ($B-V$), where $\lambda$ is one of the 
wavelengths of the broad-band filters ($R,I,J,H,K,L$) provide an effective method for separating the influence of the normal extinction produced by the diffuse interstellar medium from that of the abnormal extinction arising within regions having a peculiar distribution of dust sizes (cf. Chini \& Wargau 1990, Pandey et al. 2000).

The TCDs for the cluster's region ($r<r_{cl}$) are shown in Fig. 11, where NIR data have been taken from the 2MASS. The ${E(V-\lambda)}\over {E(B-V)}$ values in the cluster region are estimated as described by Pandey et al. (2003).
The slope of the distributions $m_{cluster}$ (cf. Pandey et al. 2003) is found to be $1.21\pm0.01, 2.00\pm0.03,2.39\pm0.04, 2.47\pm0.05$ for $(V-I), (V-J), (V-H), (V-K)$ vs $(B-V)$ diagrams respectively. The ratio of total-to-selective extinction in the cluster region, $R_{cluster}$, is derived using the procedure given by Pandey et al. (2003). 
The ratios ${E(V-\lambda)}\over {E(B-V)}$  $(\lambda \ge \lambda_I)$ yield $R_{cluster} = 3.06 \pm 0.07$ which indicates a normal grain size in the cluster region.
Tapia et al. (1991) also concluded that the interstellar extinction law in the direction of this cluster appears to be similar to the average Galactic law. Recently Negueruela et al. (2007) have also supported a normal reddening law in the cluster region.

\begin{figure}
\centering
\includegraphics[height=8.5cm,width=7cm,angle=-90]{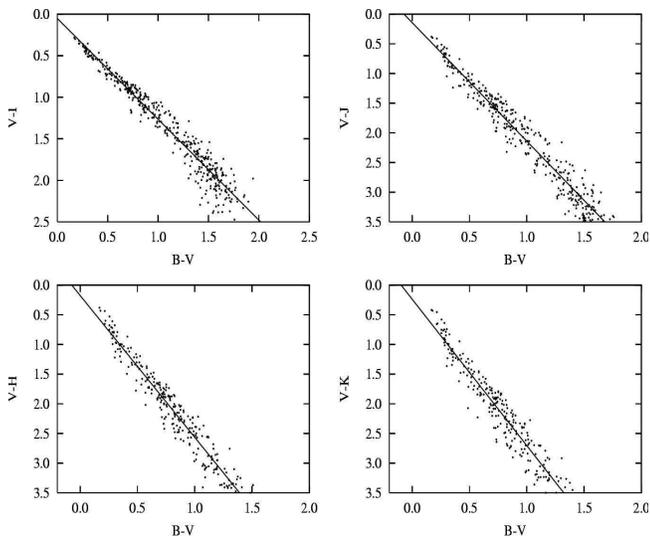}
\caption{$(V-I), (V-J), (V-H), (V-K)$ vs $(B-V)$ two color diagram for stars within cluster region 
($r<r_{cl}$). Straight line shows least square fit to the data. }
\end{figure}

\section{Optical Color Magnitude Diagrams}

\subsection {Distance and age of the cluster}

\begin{figure*}
\centering
\includegraphics[height=16cm,width=18cm]{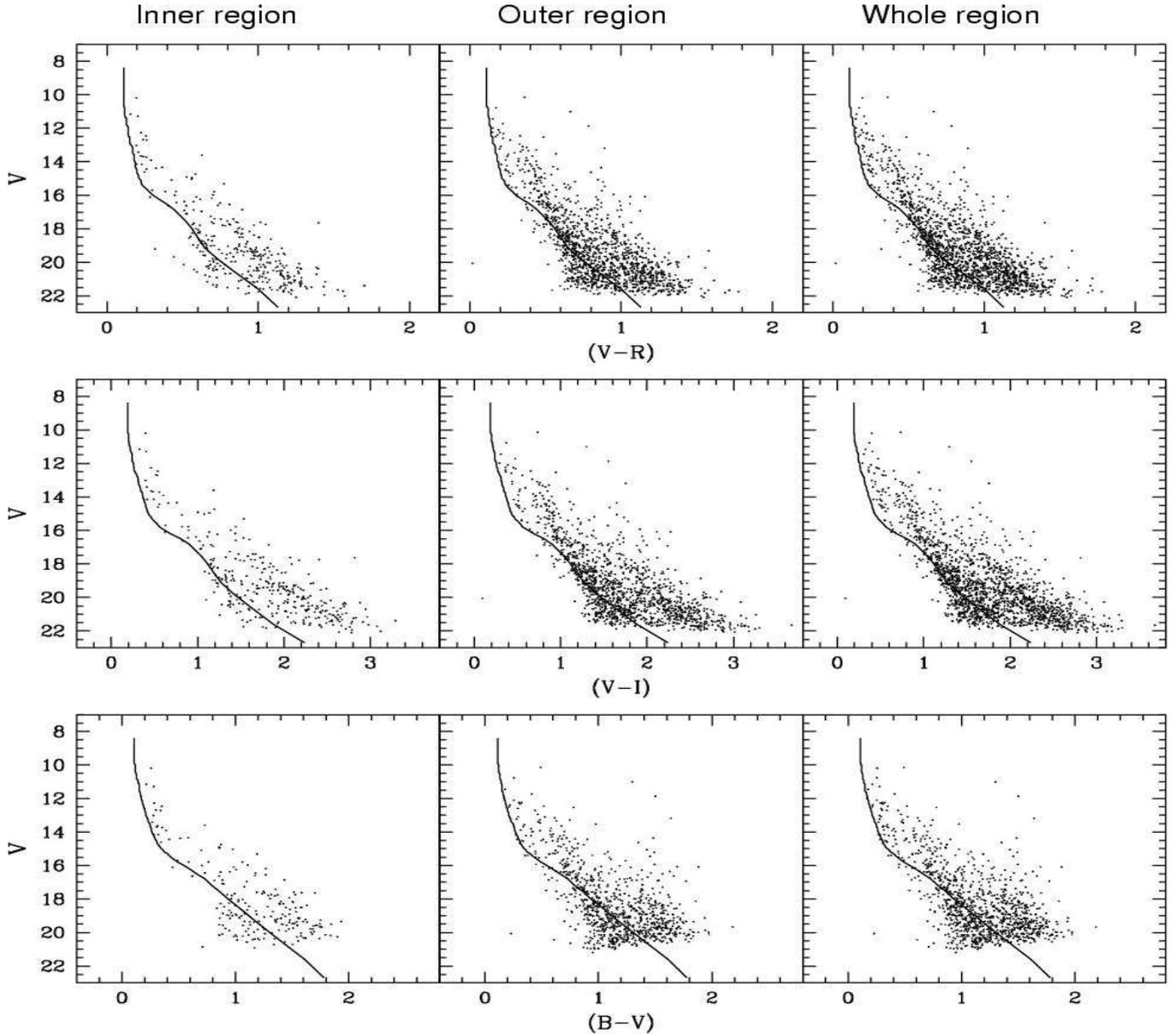}
\caption{The CMDs for stars lying in the two subregions and whole cluster region. 
The isochrone by Bertteli et al. (1994) for solar metallicity and age = 4 Myr are also plotted for $E(B-V)_{min}=0.4$ and distance modulus of 13.8 mag. }
\end{figure*}

Colour-magnitude diagrams (CMDs) for stars in two subregions and for whole cluster region are shown in Fig. 12.  A broad MS,  due to variable reddening in the cluster region, extending up to $V \sim$ 15 mag can be noticed in the inner region of the cluster. Distribution of stars fainter than  $V \sim$ 15 mag deviates towards red side of MS indicating presence of PMS stars in the cluster region. The effect of field star contamination increases as we move towards outer region of the cluster. 

Using $E(B-V)_{min} =0.40$ mag and following relations $E(U-B)/E(B-V)$=0.72, $A_V=3.1\times E(B-V)$, $E(V-R)=0.60\times E(B-V)$, $E(V-I)=1.25\times E(B-V)$, we visually fit theoretical isochrone for log age = 6.6 (4 Myr) and  $Z=0.02$ by Bertelli et al. (1994) to the blue envelope of the observed MS and found a distance modulus $(m-M)_V = 13.8 \pm 0.15$ corresponding to a distance of $3.25 \pm 0.20$ kpc. The reported values of the distance for the cluster NGC 1893 
lie in the range of 3.6 to 4.3 kpc barring the distance (6 kpc) reported by Marco et al. (2001).

\subsection {Probable members of the cluster}

To study the LF/MF, it is necessary to remove field star contamination from the sample of stars in the cluster region. Membership determination is also crucial for assessing the presence of PMS stars, because PMS stars and dwarf foreground stars both occupy similar positions above the zero-age main sequence (ZAMS) in the CMD. In the absence of  proper motion study, we used statistical criterion to estimate the number of probable member stars in the cluster region. 

Figs 13 a and b show $V, (V-I)$ CMDs for the cluster region and field region respectively. The contamination due 
to background field population is clearly visible in the cluster region CMD. The background population may belong to Norma-Cygnus (outer) arm (cf. Carraro et al. 2005, Pandey et al. 2006). 
The thick line on the cluster CMDs demarcates the contamination due to background population.
The stars to the left of the line  in the cluster CMD are considered as field stars and have not been further considered in the analysis. The cluster CMD shows a significant number of stars towards the right of the line. Majority of these stars should be PMS stars. To remove contamination of field stars from the MS and PMS sample, we statistically subtracted the contribution of field stars from the CMD of the  cluster region using the following procedure. For a randomly selected star in the $V, (V-I)$ CMD of the field region, the nearest star in the cluster's $V,(V-I)$ CMD within $V\pm0.125$ and $(V-I)\pm0.065$ of the field star was removed. While removing stars from the cluster CMD, necessary corrections for incompleteness of the data samples were taken into account.  The statistically cleaned $V, (V-I)$ CMD of the cluster region is shown in Fig. 13 c which clearly shows the presence of PMS stars in the cluster. 

\begin{figure}
\centering
\includegraphics[height=6cm,width=8cm]{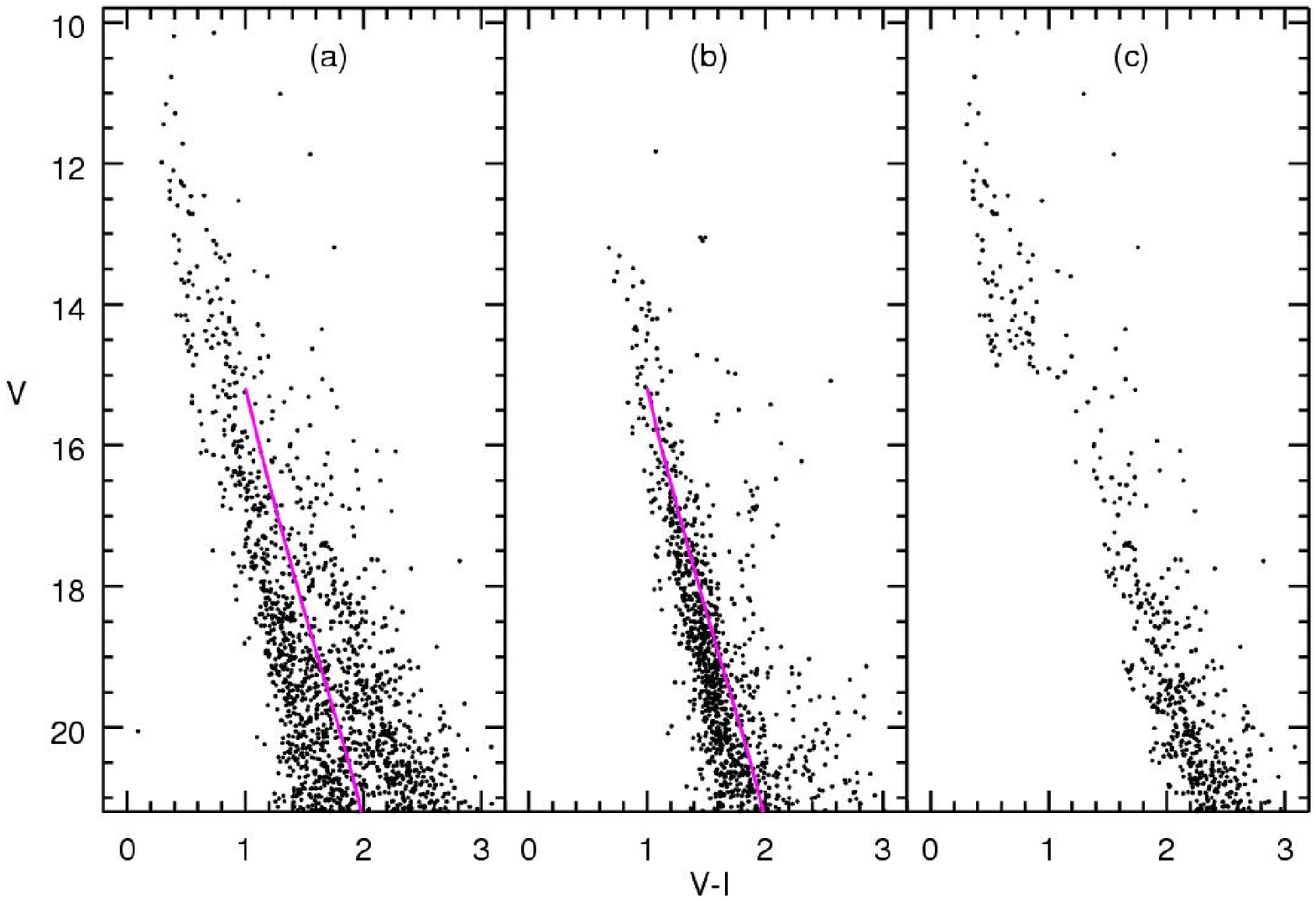}
\caption{$V/(V-I)$ CMD for (a) stars in the cluster region, (b) stars in the field region and (c) statistically cleaned CMD. Thick line demarcates the contribution background field star contamination.}
\end{figure}

Fig. 14 shows statistically cleaned unreddened $V_0/(V-I)_0$ CMD where stars having spectral type earlier than $A0$ were individually unreddened (cf. \S 7.1), whereas mean reddening of the region, estimated from available individual reddening values in that region, was used for other stars. In Fig. 14 we have plotted the isochrone for 4 Myr by Bertelli (1994) and PMS isochrones by Siess et al. (2000). Evolutionary tracks by Siess et al. (2000) for various masses are also plotted. Fig. 14 manifests that PMS population has an age spread of about 1-5 Myr. To check the reality of the age spread of PMS population, we plotted  $V/(V-I)$ CMD for $H\alpha$ emission stars and NIR excess stars (see \S 9.1) in Fig. 15, which also indicates an age spread of about 1-5 Myr for the probable PMS stars.

\begin{figure}
\centering
\includegraphics[height=8cm,width=7cm]{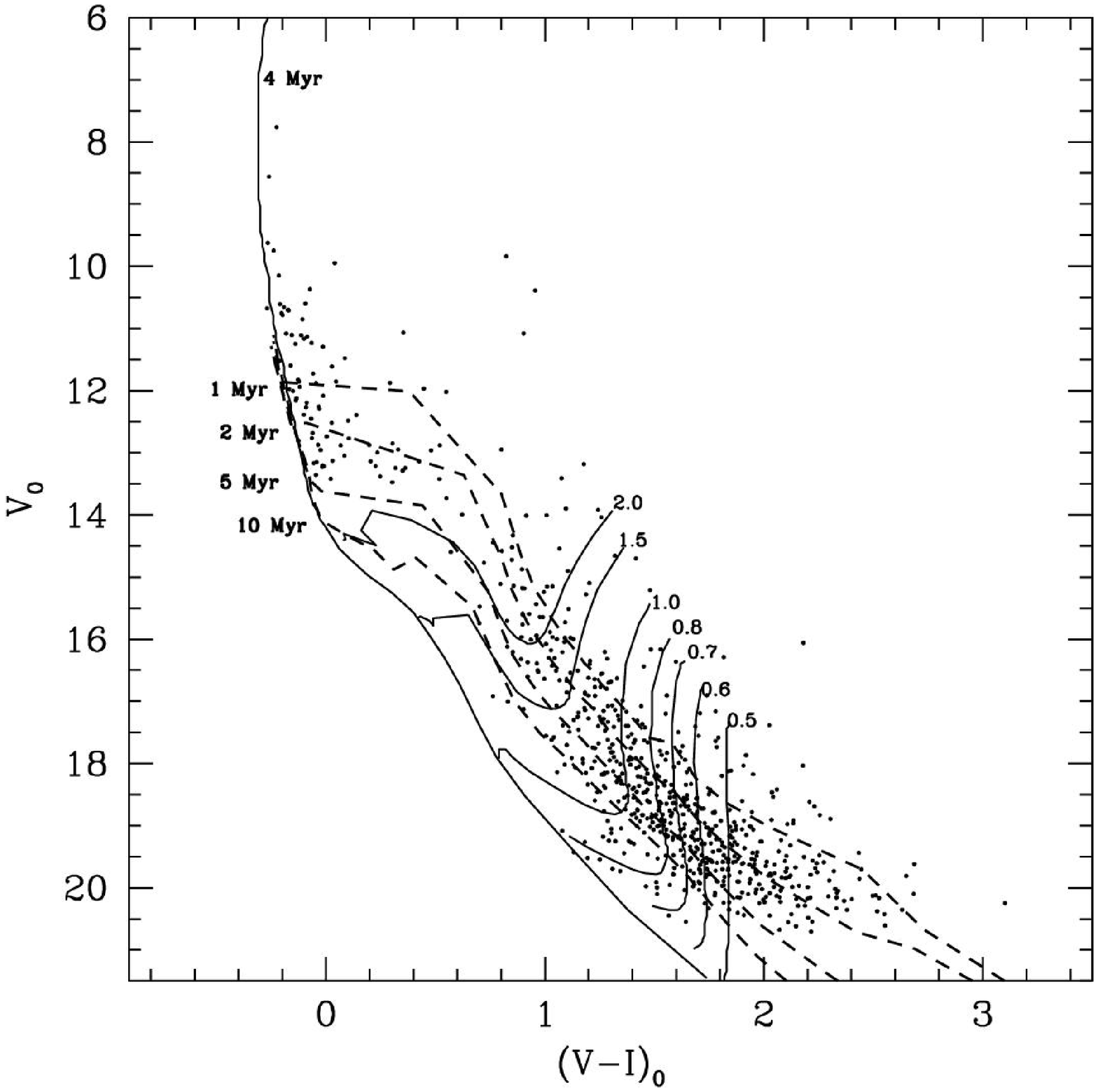}
\caption{Statistically cleaned $V_0/(V-I)_0$ CMD for stars lying in the cluster region. 
The isochrone for 4 Myr age by Bertelli et al. (1994) and PMS isochrones 
of 1,2,5,10 Myr along with evolutionary tracks of different mass stars by Siess et al. (2000) are also shown. All the isochrones are corrected for a distance of 3.25 kpc.}
\end{figure}

\begin{figure}
\centering
\includegraphics[height=8cm,width=8cm]{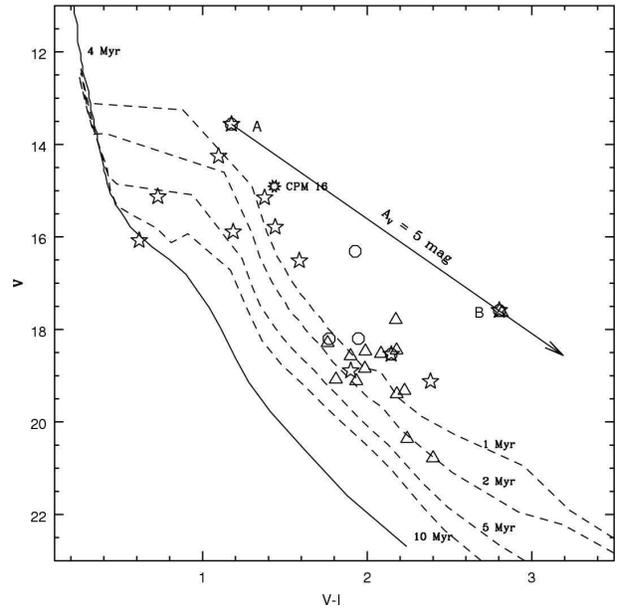}
\caption{$V,(V-I)$ CMD for $H\alpha$ emission stars (stars and open circles, cf. Sec 9.1) and NIR excess stars (triangles, cf. Sec 9.1). Isochrone for 4 Myr by Bertelli et al. (1994, continuous line) and PMS isochrones for 1,2,5,10 Myr by Siess et al. (2000) (dashed lines) are also shown. All the isochrones are corrected for distance of 3.25 kpc and reddening $E(B-V)=0.4$. CPM16 and probable HBe stars (cf. Sect 9.1) are also marked by their ID numbers as given in Table 3. The straight line indicates reddening vector for A$_V$= 5 mag. }
\end{figure}

\section{Near Infrared colour-colour and colour-magnitude diagrams}

NIR observations are very effective for investigating the nature of obscured clusters and the populations of YSOs which are embedded in molecular clouds (Lada \& Adams 1992).

\subsection{Colour-colour diagram}

In Fig. 16, we have plotted $(J-H)/(H-K)$ colour-colour diagram for the cluster region ($r<r_{cl}$) and a nearby reference field. The solid and thick dashed curves represent the unreddened main-sequence and giant branch (Bessell \& Brett 1988). The parallel dashed lines are the reddening vectors for early- and
late-type stars (drawn from the base and tip of the two branches). The dotted line indicates the locus of T Tauri stars (Meyer et al. 1997). The extinction ratio $A_J/A_V = 0.265, A_H/A_V = 0.155$ and $A_{K_s}/A_V=0.090$ have been taken from Cohen et al. (1981). 
All the 2MASS magnitudes and colours as well as the curves are in the CIT system. We classified sources into three regions in the CC diagram (cf. Ojha et al. 2004a and references therein). `F' sources are located between the reddening vectors projected from the intrinsic color of main-sequence stars and giants and are considered to be field stars (main-sequence stars, giants) or Class III /Class II sources with small NIR excesses. `T' sources are located redward of region `F' but blueward of the reddening line projected from the red end of the T Tauri locus of Meyer et al. (1997). These sources are considered to be mostly classical T Tauri stars (Class II objects)
with large NIR excesses. There may be an overlap in NIR colours of Herbig Ae/ Be stars and T Tauri stars in the `T' region (Hillenbrand et al. 1992). `P' sources are those located in the region redward of region `T' and are most likely Class I objects (protostar-like objects; Ojha et al. 2004b).
Comparison of Fig. 16 (a) and (b) shows that the stars in cluster region are distributed in a much wider range as compared to those in reference field. A significant number of sources in the cluster region exhibit NIR excess emission, a characteristics of young stars with circumstellar material. Fig. 16 (a) also shows a distribution of $H\alpha$ emission stars. 
$H\alpha$ emission stars detected from slitless spectroscopy (star symbols) are distributed in Herbig Ae/Be region as well as in `F' region. whereas $H\alpha$ emission stars from narrow-band photometry (open circles) are distributed only in the region `F' of the colour-colour diagram.

The $H\alpha$ emission star B (Table 3) is relatively more extincted and shows a large NIR excess. This star lies near to $H\alpha$ emission star A, which has been assigned as an HBe star (S3R1N3) by Marco \& Negueruela, 2002. The $V, (V-I)$ CMD (Fig. 15) indicates that the star B is more extincted by $\sim 4.0$ mag in comparison to star A. A correction of $A_V \sim 4.0$ mag puts star B in the HAeBe region in Fig. 16.

\begin{figure*}
\centering
\hbox{
\includegraphics[height=8.2cm,width=9cm]{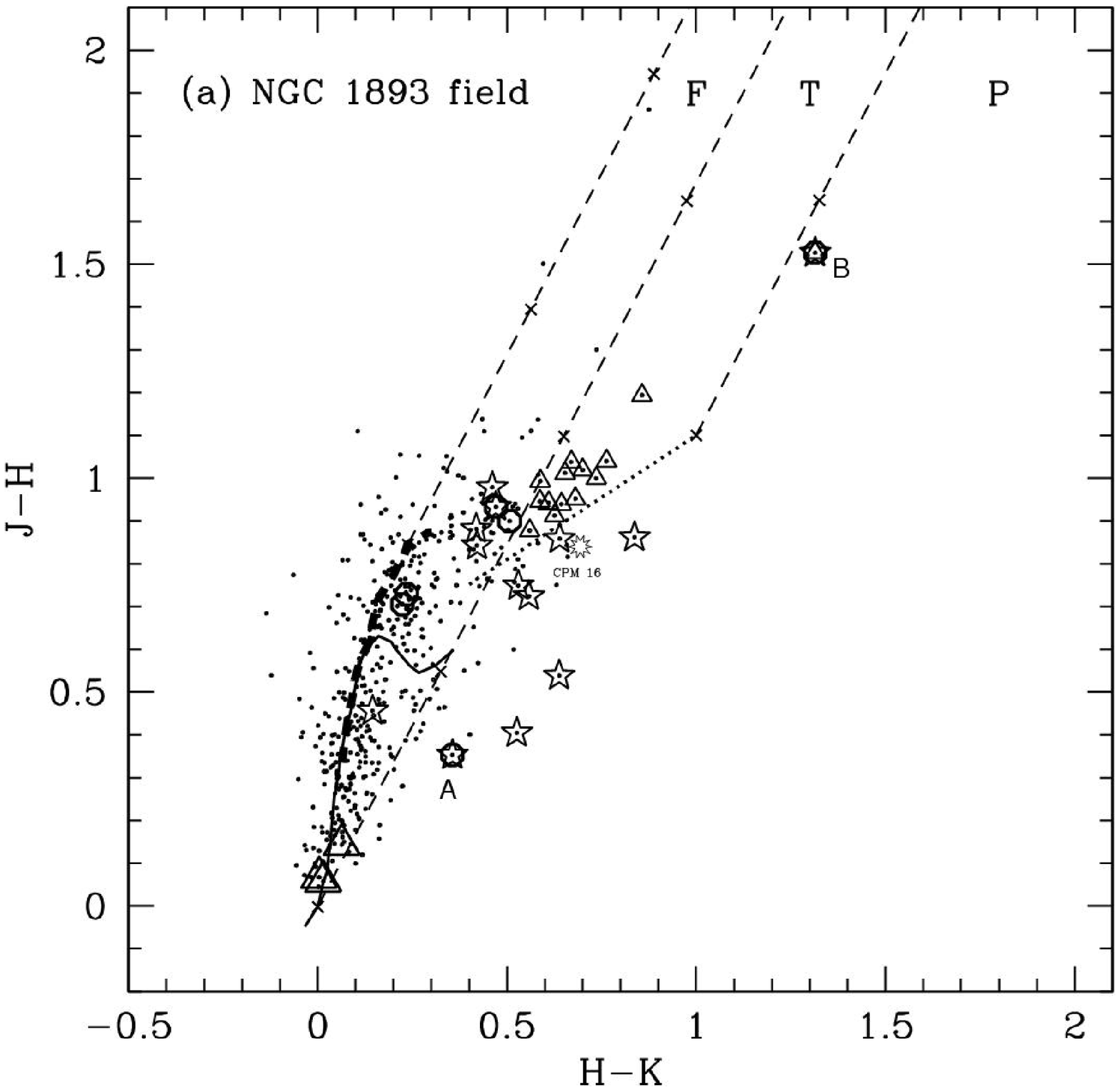}
\includegraphics[height=8.2cm,width=9cm]{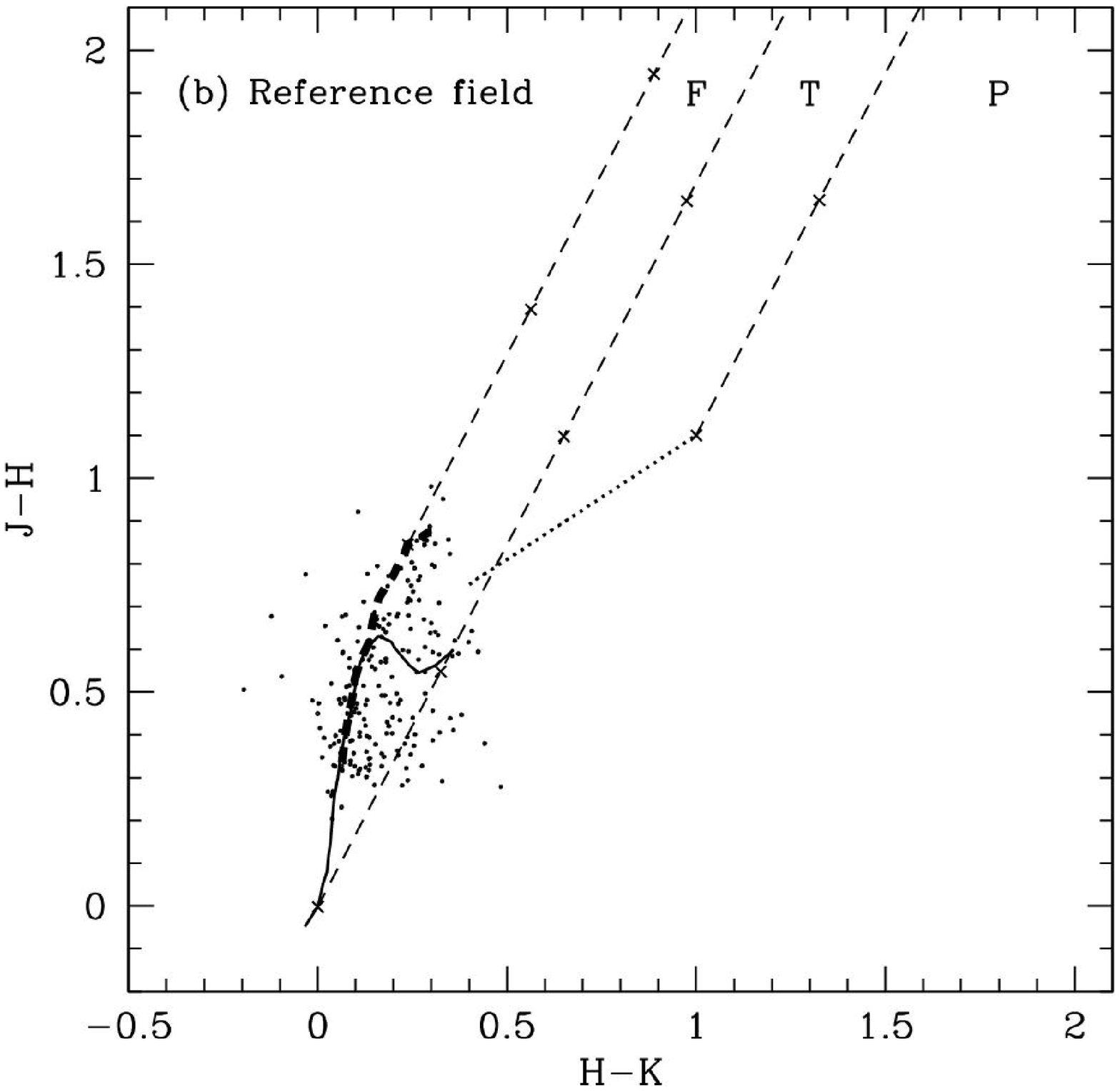}
}
\caption{$(J-H)/(H-K)$ colour-colour diagrams of sources detected in the $JHK_s$ bands with photometric errors less than 0.1 mag in (a) the cluster region ($r\le r_{cl}$) and (b) the reference field. The sequences for dwarfs (solid curve) and giants (thick dashed curve) are from Bessell \& Brett (1988). The dotted line represents the locus of T Tauri stars (Meyer et al. 1997). Dashed straight lines represent the reddening vectors (Cohen et al. 1981). The crosses on the dashed lines are separated by $A_V$ = 5 mag. The star symbols and open circles are the $H\alpha$ emission stars detected from slitless spectroscopy and narrow band photometry respectively. Open triangles are probable YSOs (T-Tauri stars). Large open triangles are O-type stars. Probable HBe stars and CPM16 are marked.} 

\end{figure*}

\begin{figure*}
\centering
\hbox{
\includegraphics[height=8.2cm,width=9cm]{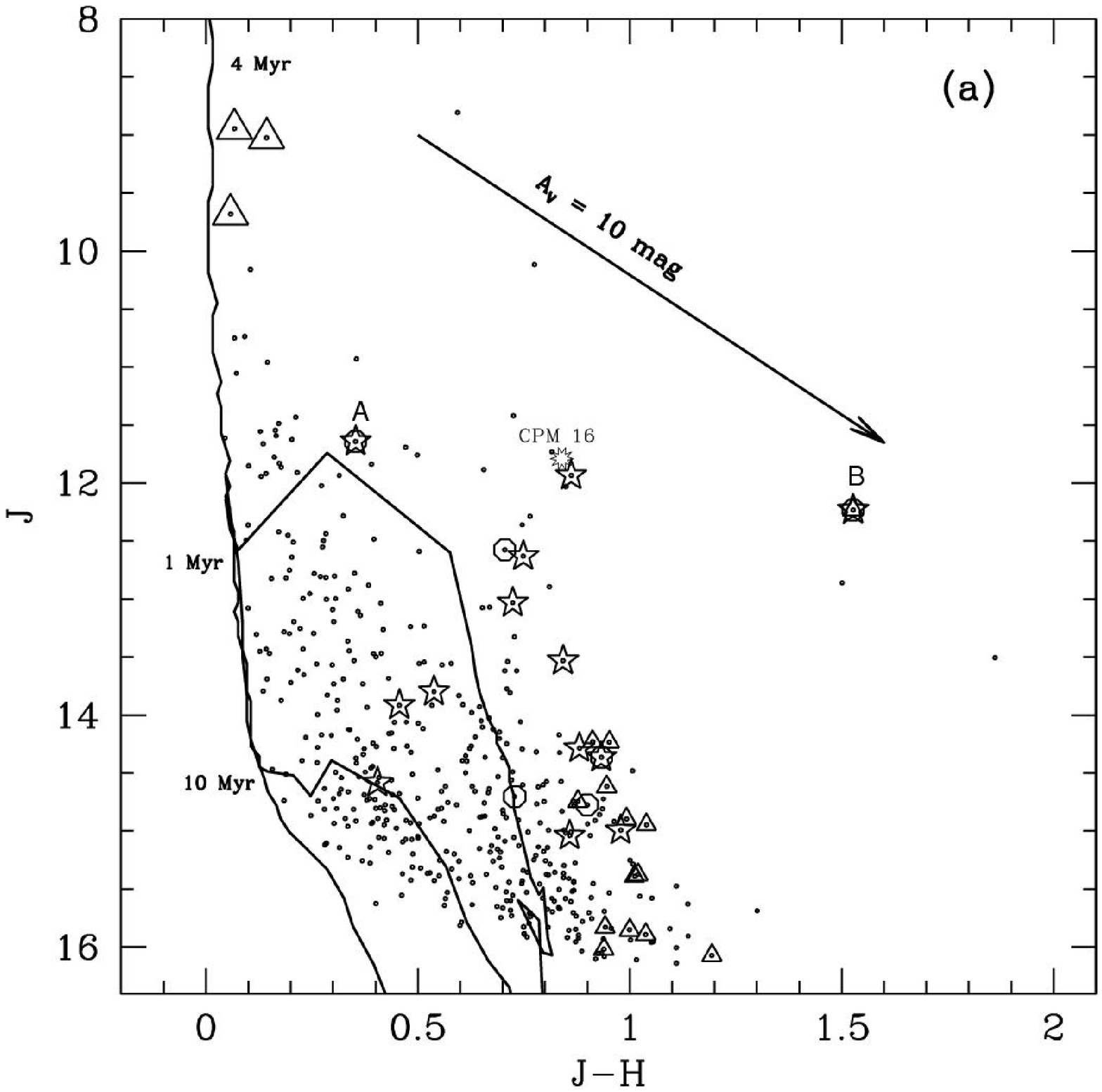}
\includegraphics[height=8.2cm,width=9cm]{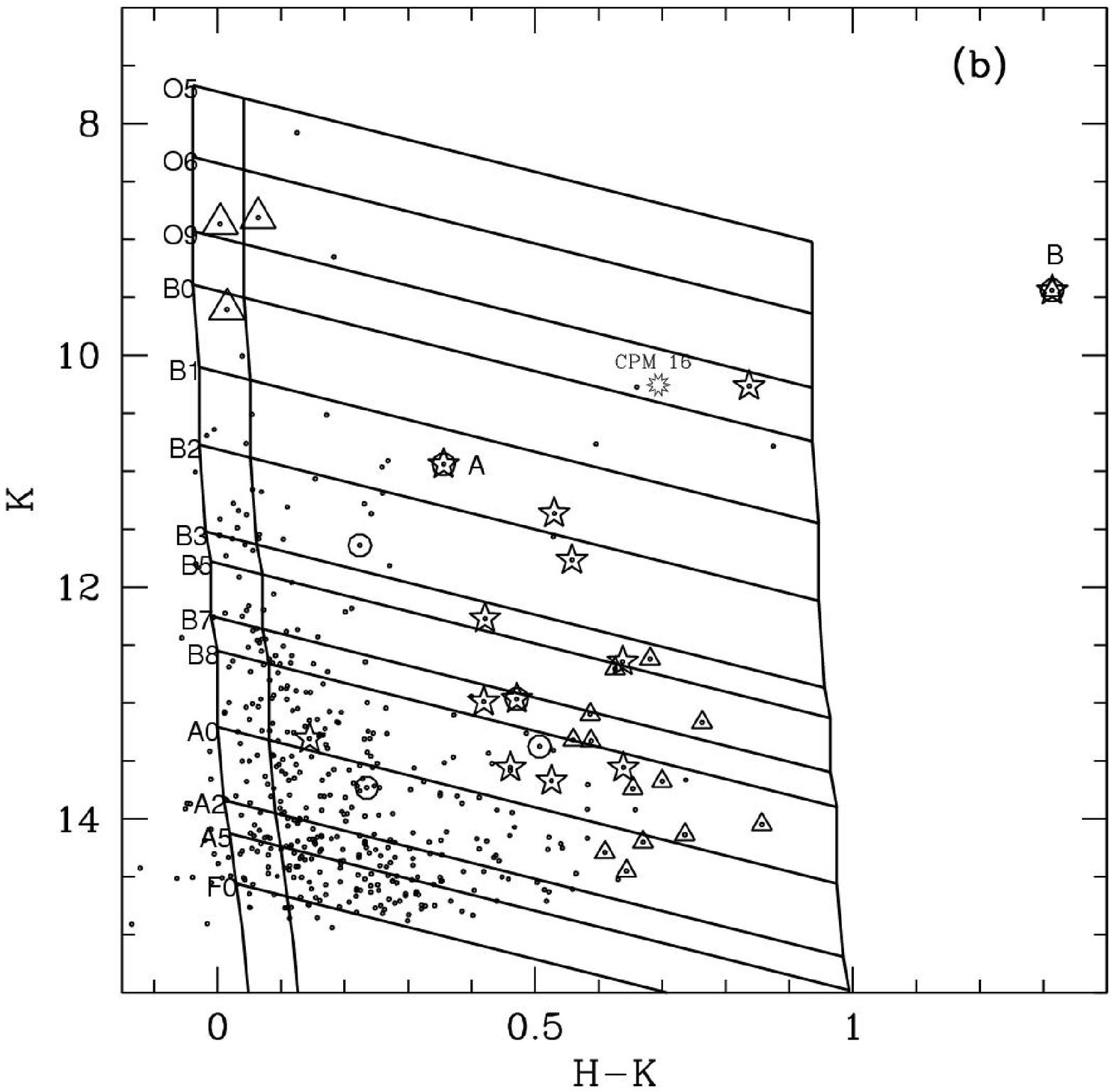}
}
\caption{(a) $J/(J-H)$ CM diagram of the stars within cluster region ($r\le r_{cl}$).
The symbols are same as in Fig. 16. The isochrone of 4 Myr (Z=0.02)  by Bertelli et al. (1994) and PMS 
isochrones of age 1 and 10 Myr taken from Siess et al. (2000) corrected for a distance of 3.25 kpc and 
reddening $E(B-V)_{min}=0.40$ are also shown.
(b) $K/(H-K)$ CM diagram for the stars within cluster region ($r\le r_{cl}$).
The symbols are same as in Fig. 16. The vertical solid lines from left to right indicate the main-sequence track at 3.25 kpc reddened by $A_V=0,1.24,15$ mag respectively. The intrinsic colors are taken from Koorneef (1983). Slanting horizontal lines 
identify the reddening vectors for each spectral type marked in the figure. 
Probable HBe stars and CPM16 are marked.}
\end{figure*}

\subsection{Colour-magnitude diagram}

The CMDs are useful tool for studying the nature of the stellar population within star-forming regions. In Fig. 17 (a), we have plotted $J/(J-H)$ colour magnitude diagram for the stars within cluster region. Using the relation ${{A_J}\over {A_V}}=0.265$, ${{A_{J-H}}\over{ A_V}}=0.110$ (Cohen et al. 1981) and $A_V=3.1\times E(B-V)$, the isochrones for age 4 Myr and PMS isochrones for ages 1, 10 Myr by Bertelli et al. (1994) and Siess et al. (2000) respectively have been plotted assuming the distance of 3.25 kpc and extinction $E(B-V)_{min}=0.4$ as obtained from the optical data. Probable YSOs (T-Tauri type stars) obtained from the colour-colour diagram and $H\alpha$ emission stars, are shown by open triangles and star symbols/open circles respectively. Most of the probable T Tauri and $H\alpha$ emission stars have age $\le$ 1 Myr.
Fig. 17 (b) shows $K/(H-K)$ CMD for the stars lying within the cluster region. The symbols are the same as in Fig 16. 
The nearly vertical lines represent the ZAMS at a distance of 3.25 kpc reddened by $A_V$ = 0,  1.24 and 15 mag respectively. The intrinsic colours are taken from Koorneef (1983). The parallel slanted lines trace the reddening zones for each spectral type. 

Fig. 18 represents $J/(J-H)$  diagram for probable YSO candidates identified in Fig 16. The symbols are same as in Fig 16.  The mass of the  probable YSO candidates can be estimated by comparing their location on the CMD with the evolutionary models of PMS stars. The solid curve, taken from Siess et al. (2000) denotes the locus of 1 Myr old PMS stars having masses in the range of 0.1 to 3.5 $M_\odot$. To estimate the stellar masses, the $J$ luminosity is recommended rather than that of $H$ or $K$, as the $J$ band is less affected by the emission from circumstellar material (Bertout et al. 1988). The majority of the YSOs have masses in the range 3.0 to 1.0 $M_\odot$ indicating that these may be 
T Tauri stars. A few stars having mass greater that  3.0 $M_\odot$ may be candidates for Herbig Ae/Be stars.

\section{Initial mass function}

The distribution of stellar masses that form in a star-formation event
in a given volume of space is called Initial Mass Function (IMF) and
together with star formation rate, the IMF dictates the evolution and fate
of galaxies and star clusters (Kroupa 2002).

      The mass function (MF) is often expressed by the power law,
 $N (\log m) \propto m^{\Gamma}$ and  the slope of the MF is given as:

    $$ \Gamma = d \log N (\log m)/d \log m  $$

\noindent

 where $N  (\log m)$ is the number of star per unit logarithmic mass interval. In the solar neighborhood the classical value derived by Salpeter  (1955) is $\Gamma = -1.35$. 

With the help of statistically cleaned CMD, shown in Fig. 14,  we can derive the 
MF using the theoretical evolutionary model of Bertelli et al. (1994). Since 
post-main-sequence age of the cluster is $\sim$ 4 Myr, the stars having $V<15$ mag 
($V_0 < 13.5; M>2.5 M_\odot$) have been considered to be on the main sequence. For the MS 
stars, the LF was converted to the MF using the theoretical model by Bertelli et al. (1994) (cf. Pandey et al. 2001, 2005). The MF for PMS stars was obtained by counting the number of stars in various mass bins (shown as evolutionary tracks) in Fig. 14. The MF in two subregions as well as for the whole cluster is given in Table 7 and plotted in Fig. 19.

Table 8 indicates that the slopes of the MF of MS and of PMS stars in the inner
region of the cluster are almost the same and can be represented by 
$\Gamma = -1.19\pm0.03$ which is slightly shallower than the Salpeter value.
This indicates that there is no mass segregation in the inner region of the 
cluster. In the outer region as well as for the whole cluster region, the $\Gamma$
for PMS stars seems to be shallower than the $\Gamma$  for the MS stars indicating
a break in the slope of the MF at $\sim 2 M_\odot$. However a single slope for 
the MF, in the outer region and whole cluster, can also be fitted to the entire 
observed mass range ( $0.6<M/M_\odot < 17.7$) with $\Gamma = -1.32\pm0.12$ and 
$\Gamma = -1.27\pm0.08$ respectively. 

Effect of mass segregation can be seen on the MS stars. The $\Gamma$ for the MS stars is steeper in the outer region indicating a preferential distribution of relatively massive stars towards the cluster center, whereas  $\Gamma$ for the PMS stars is steeper in the inner region as compared to the $\Gamma$ in the outer region. However the difference in the slopes is rather small ($2.5 \sigma$ and $2.7 \sigma$ in the case of MS and PMS stars respectively).

For the mass range $2.5\le M/M_\odot \le 17.7$, the MF for the whole cluster 
region ($r\le6'$) can be represented by $\Gamma = -1.71\pm0.20$ which is in agreement
with the value ($\Gamma = -1.6\pm0.3$) given by Massey et al. (1995a). 

 %------------------------------------------------------------------------------
\begin{table*}
\centering
\begin{minipage}{140mm}
\caption{The mass function of the NGC 1893 cluster. The number of probable cluster
members ($N$) have been obtained after subtracting the expected contribution of
field stars. log $\phi$ represents log ($N$/dlog $m$).}
\begin{tabular}{@{}rrrrrrrr@{}}
\hline

Mass&Mean&\multicolumn{2}{c}{Inner region}&\multicolumn{2}{c}{Outer region}&\multicolumn{2}{c}{Whole
region}\\
$(M/M_{\odot})$&log $M/M_{\odot}$& N&log $\phi$& N& log $\phi$&N& log $\phi$ \\ \hline

&&&&&&&\\
  17.7  - 12.3  & 1.1759&   1  & 0.7964  &     1 & 0.7964 &     2  &1.0974   \\
  12.3  -  8.4  & 1.0147&   2  & 1.0882  &     3 & 1.2643 &     5  &1.4861   \\
   8.4  -  5.6  & 0.8451&   4  & 1.3490  &    10 & 1.7470 &    14  &1.8931   \\
   5.6  -  3.7  & 0.6650&   6  & 1.5191  &    16 & 1.9451 &    22  &2.0834   \\
   3.7  -  2.5  & 0.4868&   8  & 1.6644  &    25 & 2.1592 &    33  &2.2798   \\
   2.0  -  1.5  & 0.2432&   13 & 2.0173  &     37&  2.4715&      50&    2.6023\\
   1.5  -  1.0  & 0.0969&   24 & 2.1345  &     57&  2.5101&      81&    2.6628\\
   1.0  -  0.8  &-0.0459&   20 & 2.3147  &     46&  2.6764&      66&    2.8332\\
   0.8  -  0.7  &-0.1249&   17 & 2.4671  &     31&  2.7280&      48&    2.9179\\
   0.7  -  0.6  &-0.1872&   20 & 2.4753  &     40&  2.7763&      60&    2.9527\\

\hline
\end{tabular}
\end{minipage}
\end{table*}

\begin{figure}
\centering
\includegraphics[height=8cm,width=8cm]{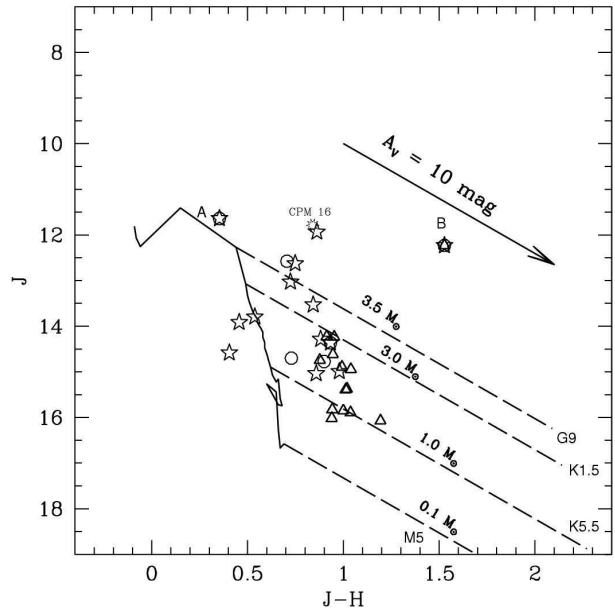}
\caption{$J/(J-H)$ CMD for the YSO candidates in cluster region ($r\le r_{cl}$).
The symbols are same as in Fig. 15. The solid curve denotes PMS isochrone of 1 Myr,
derived from Siess et al. (2000), corrected for a distance of 3.25 kpc.  
Masses range from 3.5 to 0.1 $M_\odot$
from top to bottom. The dashed oblique reddening lines denote the positions of PMS 
stars of 0.1, 1, 3.0 and 3.5 $M_\odot$. }
\end{figure}

\begin{figure}
\centering
\includegraphics[height=8cm,width=10cm,angle=-90]{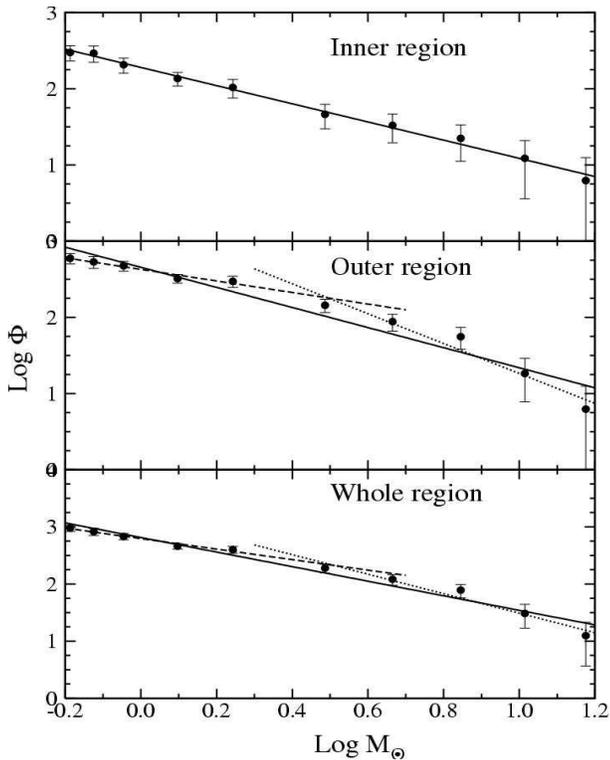}
\caption{A plot of the mass function in the cluster. log $\phi$ represents log ($N$/dlog $m$). The error bars represent $\pm \sqrt N$ errors. The solid line shows a least square fit to the entire mass range $0.6<M/M_\odot <17.7$. }
\end{figure}

\section{K-band Luminosity function}

In order to obtain the K-band Luminosity function (KLF) of the cluster, it is essential to take into account the incompleteness of the data as well as the foreground and background source contaminations. 
The completeness of the data is estimated using the ADDSTAR routine of DAOPHOT as described in Sec. 5. To take into account foreground/ background field star contamination we used both the Besan\c con Galactic model of stellar population synthesis (Robin et al. 2003) and the nearby reference field stars. Star counts are predicted using the Besan\c con model in the direction of the control field.
We checked the validity of the simulated model by comparing the model KLF with that 
of the control field (see Fig. 20 (a)) and found that both the KLFs match rather well. 
An advantage of using the model is that we can separate the foreground ($d<3.25$ kpc) and  background ($d>3.25$ kpc) field stars.
As mentioned in Section 7.1, the foreground extinction using optical data 
was found to be $A_V \sim1.24$ mag. The model simulations with $A_V$ = 1.24 mag 
and $d<3.25$ kpc gives the foreground contamination.
 
The background population ($d>3.25$ kpc) was simulated with $A_V$ = 1.86 mag in the model. We thus determined the fraction of the contaminating stars (foreground+background) over the total model counts. This fraction was used to scale the nearby reference field and subsequently the star counts of the modified control field were subtracted from the KLF of the cluster to obtain the final corrected KLF.
The KLF is expressed by the following power-law:

${{ \rm {d} N(K) } \over {\rm{d} K }} \propto 10^{\alpha K}$

where ${ \rm {d} N(K) } \over {\rm{d} K }$ is the number of stars per 0.5 magnitude
bin and $\alpha$ is the slope of the power law. Fig. 20 b shows the KLF for the cluster region which indicates a slope of $\alpha = 0.34\pm0.07$. The slope is consistent with the average value of slopes ($\alpha \sim 0.4$) for young clusters (Lada et al. 1991; Lada \& Lada 1995; Lada \& Lada 2003).

\begin{figure}
\centering
\includegraphics[height=8cm,width=4cm,angle=-90]{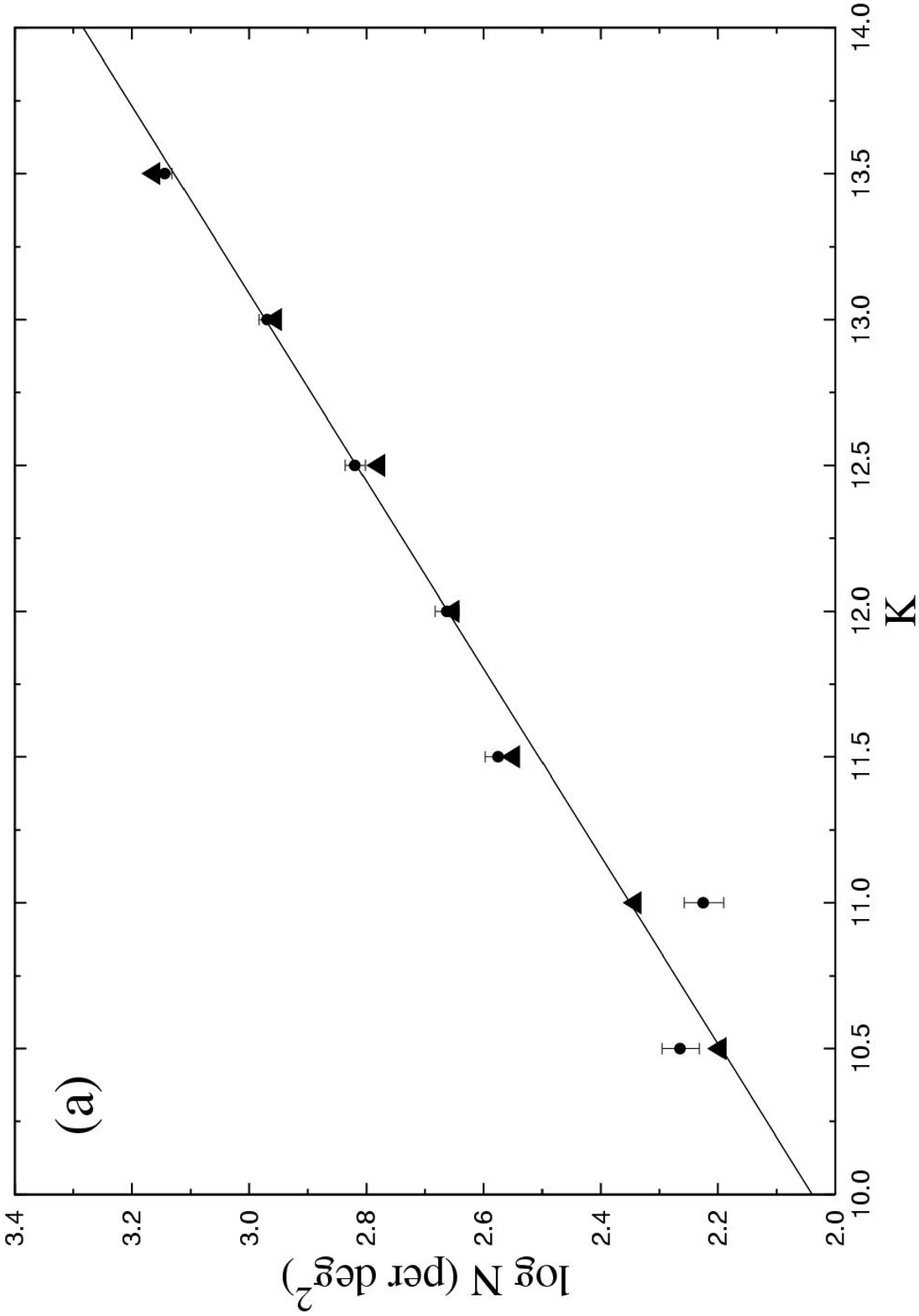}
\includegraphics[height=8cm,width=4cm,angle=-90]{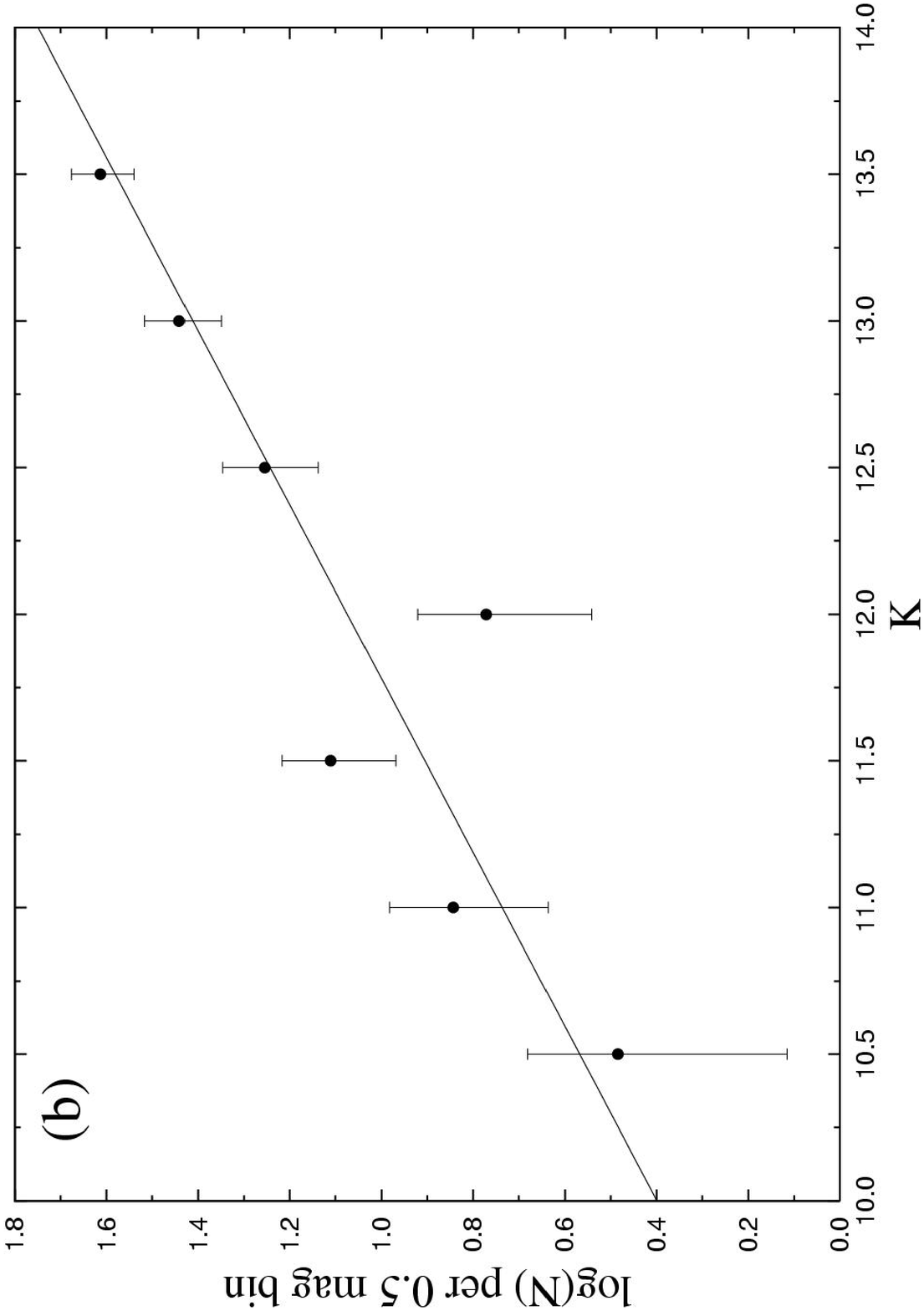}
\caption{(a) Comparison of the observed KLF in the reference field and the simulated
KLF from star count modeling. The filled circles denote the observed $K$-band  star counts in the reference field, and triangles represent the simulation from the Galactic model (see the text). The star counts are number of star per deg$^2$ and the error bars represents $\pm \sqrt N$ errors. The KLF slope ($\alpha$, see \S 11)
of the reference field (solid line) is $0.31\pm0.03$. The simulated model 
is also giving the same value of slope ($0.31\pm0.01$).
(b) The corrected KLF for the probable members in the cluster. The straight line is the
least-square fit to the data points in the mag range 10.5-13.5. }
\end{figure}

\section{Mass segregation}

There are evidences for mass segregation in a few Galactic as well as LMC star clusters, with the highest mass stars preferentially found towards the center of the cluster (see e.g. Sagar et al. 1988, Sagar \& Richtler 1991,  Pandey et al. 1992, 2001, 2005; Fischer et al. 1998).
Although, as per standard theory where stars in clusters evolve rapidly towards a state of energy equipartition through stellar encounters, consequently mass segregation - in the sense that more massive stars tend to lie near the center - is well accepted.  However, observations of mass segregation in very young clusters (e.g. Pandey, Mahara \& Sagar 1992; Hillenbrand 1997) suggest that in the case of a few young clusters the mass segregation may be the imprint of star formation process itself (Pandey et al. 2005).

\begin{figure}
\centering
\includegraphics[height=8cm,width=6cm,angle=-90]{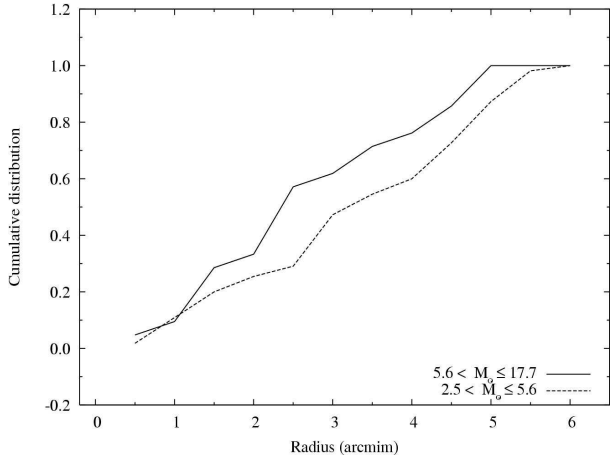}
\caption{Cumulative radial distribution of MS stars in two mass intervals.}
\end{figure}

To characterize the degree of mass segregation in  NGC 1893 we subdivided the MS sample into two mass groups  ($5.6 < M/M_{\odot} \le 17.7, 2.5 < M/M_{\odot} \le 5.6$). Fig. 21 shows cumulative distribution of MS stars as a function of radius in two different mass groups. The figure indicates an effect of mass segregation in the cumulative distribution in the sense that more massive stars ($5.6 < M/M_{\odot} \le 17.7$) tend to lie toward the  cluster center. The Kolmogorov-Smirnov test confirms the statement that the cumulative distribution of massive stars in the cluster is different from the cumulative distribution of relatively less massive stars at a confidence level of 99\%.
Because of the dynamical relaxation, low mass stars in a cluster may acquire larger random
velocities,  consequently occupy a larger volume than  high mass stars  (cf. Mathieu 1985, Mathieu \& Latham 1986,  McNamara \& Sekiguchi 1986). We estimated the relaxation time to decide whether the mass segregation discussed above is primordial or due to dynamical relaxation. To estimate the dynamical relaxation time $T_{E}$, we used the relation

\bigskip
{\large
~~~~~~~~~~~~~~~$T_{E} =  \frac{8.9\times10^5 N^{1/2} R_h^{3/2}}{\bar{m}^{1/2} log (0.4N)}$}

\bigskip

where N is the number of cluster stars,  $ R_{h}$ is the radius containing half of
the cluster mass and $\bar{m}$ is the average mass of cluster stars (Spitzer \& Hart
1971). The total number of MS stars and the total mass of the MS stars in the given
mass range ($2.5\le M/M_\odot \le 17.7$) is obtained with the help of LF/ MF. 
This mass should be considered as the lower limit of the total mass of the cluster. 
For the half mass radius we used half of the cluster extent i.e. 2.85 pc. 
Various parameters used to estimate $T_E$ for the cluster are given in Table 9. Using these numbers, the estimated relaxation time $T_{E}$ comes out to be larger than the age of the cluster. This indicates that the observed mass segregation of MS stars in the cluster should be of largely primordial nature.

\begin{table}
\centering
\caption{Mass Function $`\Gamma$'.}
\begin{tabular}{rrrr}
\hline
 Region&$`\Gamma$' (MS)               & $`\Gamma$' (PMS) & $`\Gamma$' (MS+PMS)\\
\hline
Inner  &$-1.25\pm0.13$$^{a}$ &$-1.15\pm0.11$$^{b}$&$-1.19\pm0.03$$^{c}$    \\
Outer  &$-1.96\pm0.26$$^{a}$ &$-0.76\pm0.09$$^{b}$&$-1.32\pm0.12$$^{c}$    \\
Whole  &$-1.71\pm0.20$$^{a}$ &$-0.88\pm0.09$$^{b}$&$-1.27\pm0.08$$^{c}$    \\
\hline
\end{tabular}
\label{clustable}

a:  Mass range (2.5 $<$ $M/M_{\odot}$ $<$ 17.7)\\
b:  Mass range (0.6 $<$ $M/M_{\odot}$ $<$ ~2.0)\\
c:  Mass range (0.6 $<$ $M/M_{\odot}$ $<$ 17.7)\\
\end{table}

\begin{table}
\centering
\caption{Parameters of the cluster.}
\begin{tabular}{rrrrr}
\hline
 Mass range&Total mass     & No. of & Average mass& $T_E$ \\      
           &($M_{\odot}$)&       stars      & $\bar{m}$ ($M_{\odot}$) & (Myr) \\
\hline
2.5$<M/M_{\odot}<$17.7 &$\sim$ 374&$\sim$ 78 &$\sim$ 4.8&$\sim$ 12\\
0.6$<M/M_{\odot}<$~2.0  &$\sim$ 333&$\sim$ 315&$\sim$ 1.1&$\sim$ 35 \\
0.6$<M/M_{\odot}<$17.7 &$\sim$ 710&$\sim$ 390&$\sim$ 1.8&$\sim$ 29 \\
\hline
\end{tabular}
\label{clustable}

\end{table}

\begin{figure*}
\centering
\includegraphics[height=14cm,width=16cm]{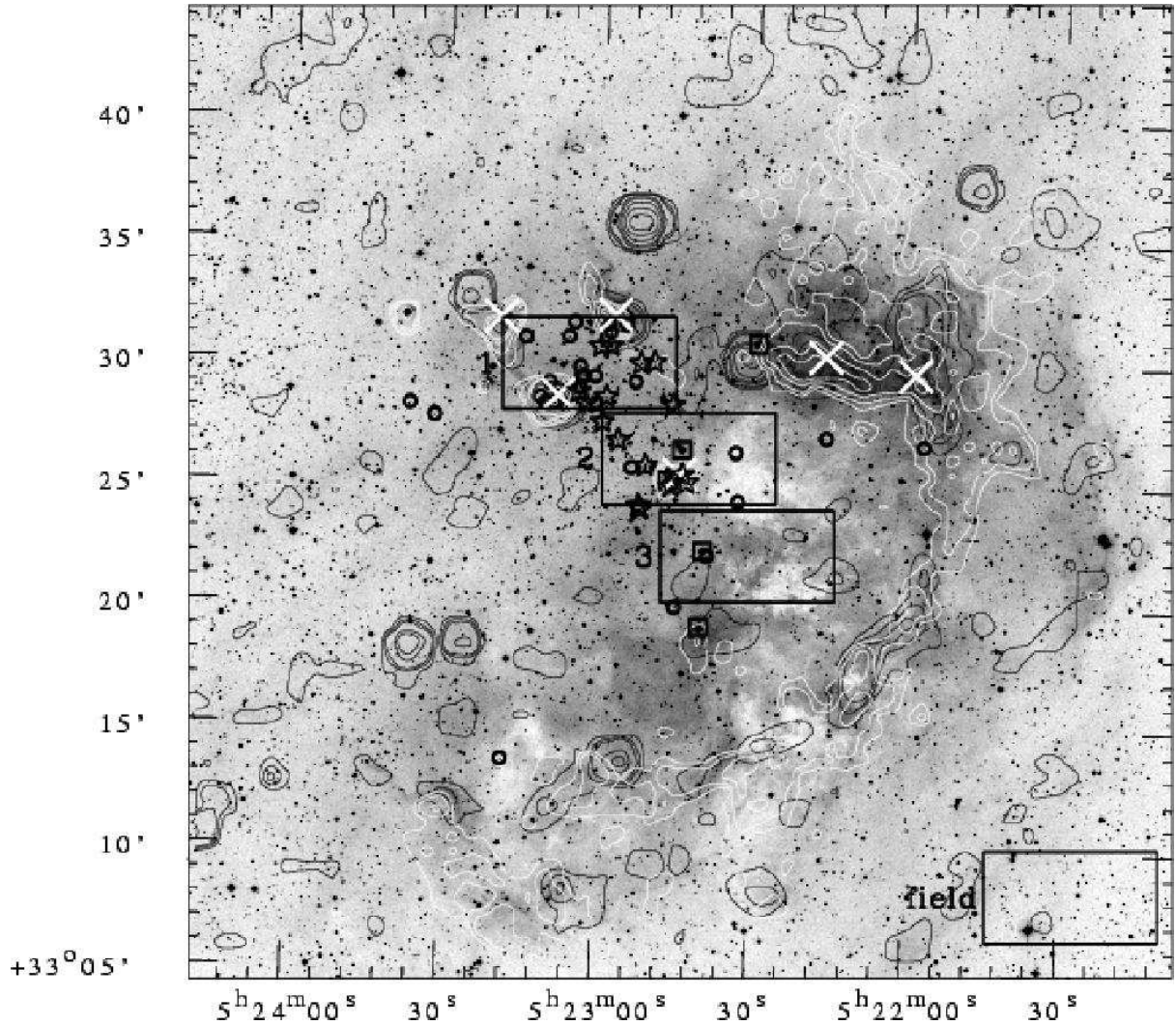}
\caption{Spatial distributions of O-type stars (open square box), IR-excess sources (probable T-Tauri type stars, 
open circles), $H\alpha$ (star symbols) stars and IRAS points sources (crosses) are 
overlaid on DSS-2 $R$ band image. NVSS ($1.4~GHz$) radio contours (${\it black ~contours}$) and MSX A-Band intensity 
contours (${\it white ~contours}$) has also been shown. 
The MSX A-Band contours are  2.5, 3, 4, 5, 10, 20, 40, 60, 80\% of the 
peak value $3.25 \times 10^{-5}$ $W m^{-2} Sr^{-1}$ and the NVSS radio contours are
.2, .5, 1, 3, 5, 10, 20, 40, 60, 80\% of the peak value 0.45 Jy/Beam.
The abscissa and the ordinates are for the J2000.0 epoch. Three subregions and a field region are also marked.}
\end{figure*}

\begin{table}
\centering
\caption{Statistics of the YSOs in the three different regions. Numbers given in parenthesis are in percentage.}
\begin{tabular}{ccccc}
\hline
Region  &Totals&NIR excess &  $H\alpha$ stars & All YSOs\\
        & stars& stars     &  (slitless)&\\
\hline
1&139     &16 (21)      &8+7$^a$ (20) & 29 (38)$^b$  \\
2&145     &5 (6)      &6+2$^a$(10) & 12 (15)$^c$\\
3&99      &1 (3)      &0 & 1 (3)\\
Field & 63& & & \\
\hline
\end{tabular}
\label{clustable}
\noindent

a: $H\alpha$ stars taken from Negueruela \& Marco et al. (2007). \\
b: Two $H\alpha$ stars have NIR excess.\\
c: One $H\alpha$ stars has NIR excess.\\
\end{table}

\begin{figure*}
\centering
\includegraphics[height=6cm,width=16cm]{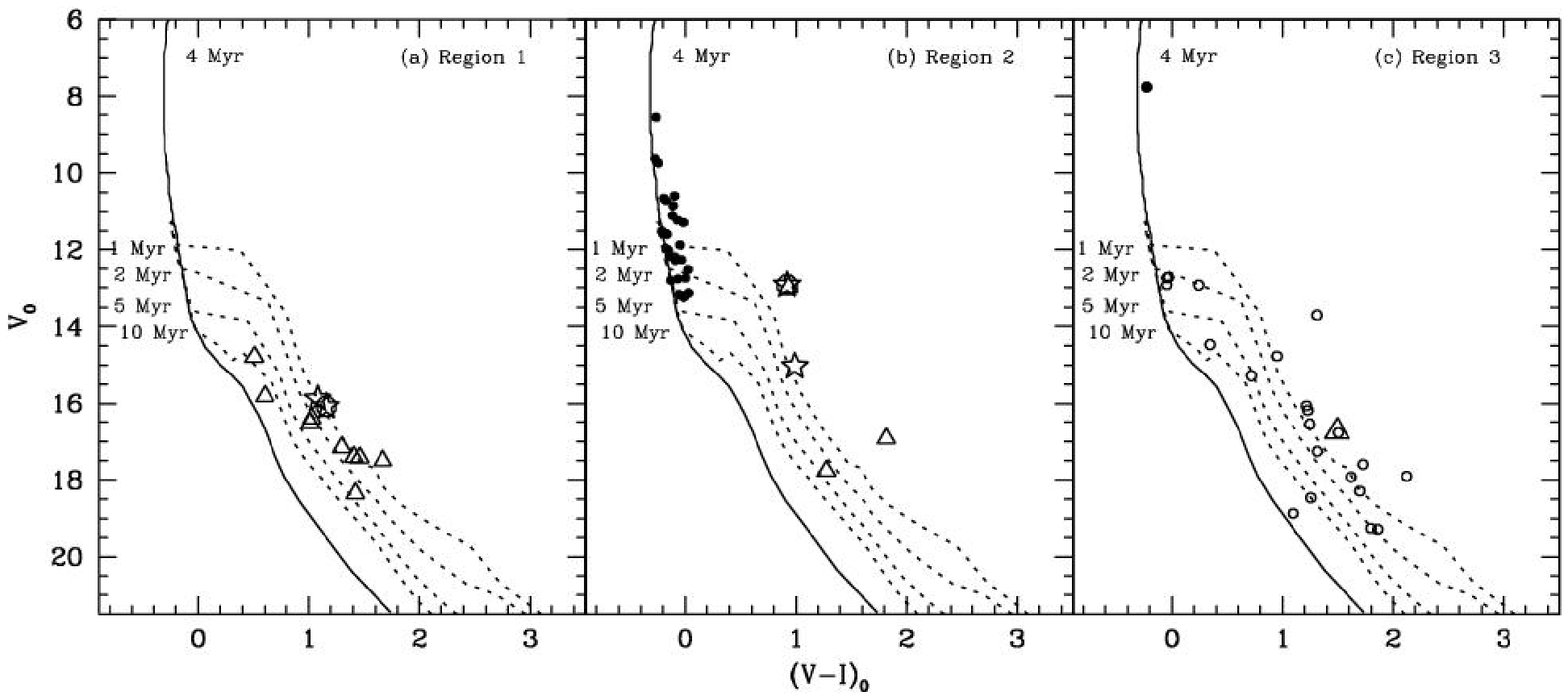}
\caption{$V_0/(V-I)_0$ CMDs for the probable YSOs in the three different regions. The triangle and star symbol represent NIR excess and $H\alpha$ stars respectively. Probable MS stars in the region 2 are shown by filled circles. Stars around O-type star in region 3 are shown by open circles.}
\end{figure*}

\section{Spatial distribution: Global overview of star formation around the cluster}

Massive O-type stars may have strong influence and significantly affect the entire star forming  regions. On one hand energetic stellar winds from massive stars could
evaporate nearby clouds and consequently terminate nearby star formation by destroying
the gas clouds. Alternatively stellar winds and shock waves from supernova explosion
may squeeze molecular clouds and induce subsequent star formation. Could both of these
effects occur in different parts of the same star-forming regions? Herbig (1962)
suggested that low and intermediate-mass stars form first and with the formation
of the most massive star in the regions, the cloud gets disrupted and star formation
ceases. Alternatively, Elmegreen \& Lada (1977) propose the low-mass stars form first
throughout the entire molecular cloud. After formation
of massive stars, the expanding ionization fronts play a constructive role to
incite a sequence of star-formation activities in the neighborhood.
The morphological details of the environment around NGC 1893 
containing several O-type stars can be used to probe the different
stages of evolution of star formation present in this cloud complex. 

In Fig. 22 spatial distribution of O-type stars, IR-excess sources (probable T-Tauri type stars) and $H\alpha$ stars have been displayed on a $40^\prime \times 40^\prime$ DSS-II $R$ band image around the cluster. Most of the YSOs seem to align toward from north-east to south-west. The apparent concentration of the YSOs can be seen around the 
nebulous region Sim 129. The distribution of YSOs indicates that star formation in the cluster might have taken place along the direction of Sim 129. 
Similar conclusion has also been reported by Negueruela \& Marco et al. (2007).
To study whether stars show any sequence in age we selected three sub-regions having equal area and a field region outside the cluster region, as 
shown in Fig. 22. Fig. 23 a and b shows  unreddened $V/(V-I)$ CMDs for the probable YSOs (triangles; NIR excess stars, star symbols; $H\alpha$ stars) in region 1 and 2 respectively. In region 2 CMD we have also plotted probable MS stars. The $(J-H)/ (H-K)$ TCD was used to estimate $A_V$ for each YSO by tracing back to the intrinsic line of Meyer et al. (1997) along the reddening vector. Probable MS stars are dereddened individually as described in $\S$ 7.1. In region 3 (Fig. 23 c) only one NIR excess star was found.  There seems to be a small clustering around O-type star in the region 3. Stars within a 1 arcmin of the O-type stars are also plotted in Fig. 23 c. These stars were unreddened using the $E(B-V)$ value of the O-type star. Fig. 23 manifests that the YSOs lying near the nebula Sim 129 have age $\sim$ 1-2 Myr. The post-main-sequence age of the cluster stars is $\sim$ 4 Myrs, whereas the YSOs near the cluster region have age $<1 - 2$ Myr. The probable PMS stars in the region 3 have age $1-5$ Myr, whereas O-type star has an age $\sim$ 4 Myr. 
To further verify the age sequence we estimate the percentage of YSOs in these three regions. The statistics is given in Table 10. $H\alpha$ photometry for the nebula Sim 129 
region was not carried out therefore $H\alpha$ photometry is not used for 
the statistics. Table 10 indicates a significant excess of YSOs in the 
region 1 (38\%) in comparison to the second (15\%) and third (3\%) region. 
The distribution of YSOs further supports a small scale sequential star formation. 
The O7-8 star (HD 242935) at the center of the cluster might have triggered the star formation towards the direction of nebula Sim 129.
%The CMDs for the three regions shown in Fig. 23 indicate that 4, 3, 4 and 2 NIR  
%excess stars have ages in the range $<$ 1 Myr, $1-2$ Myr, $2-5$ Myr and $>5$ Myr respectively, indicating that a disc half-life time of $\sim$ 2 Myr which is consistent with the disc half-life time ($\sim$ 3 Myr) suggested by Haisch, Lada \& Lada (2001).

Using the $^{12}CO$ survey Leisawitz et al. (1989) concluded that a molecular cloud 
(NGC 1893A) is associated with the cluster. The peak of the $^{12}CO$ contours is located at the SW of the cluster center. The cloud NGC 1893A is in front of the cluster and it may be receding from the cluster (Leisawitz et al. 1989). The differential extinction towards the cluster is $\sim 0.2$ mag indicating that the cloud contains low-density gas. In Fig. 22 NVSS (1.4 GHz) radio continuum emission contour maps (${\it black~ contours}$) and MSX A-band infrared contour maps (${\it white~ contours}$) are also overlaid. The location of IRAS point sources has been represented by crosses. The gross morphologies of the radio contour map and MSX A-band contour map are similar, though the emission from the MSX A-band is much more extended. The majority of emission from radio as well as MSX A-band comes from an arc-shaped region embracing the cluster from north-west to south-east. To the north-west of the cluster, the contours show relatively steep intensity gradient towards the cluster. 
This region also harbors two IRAS sources. 
%The morphology of the MSX-A band contours matches with the radio contours rather well.

\begin{figure*}
\centering
\hbox{
\includegraphics[height=7cm,width=9cm]{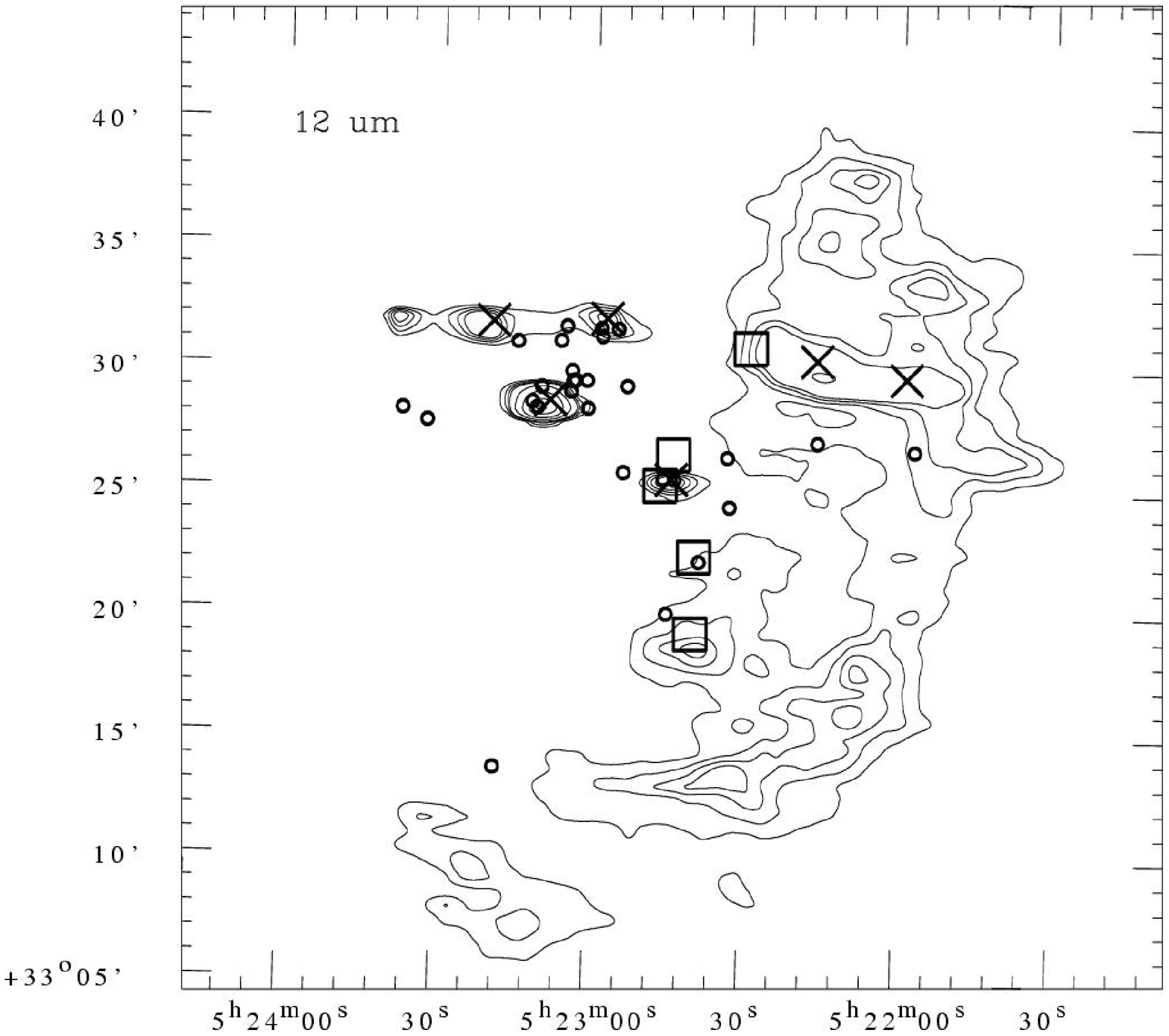}
\includegraphics[height=7cm,width=9cm]{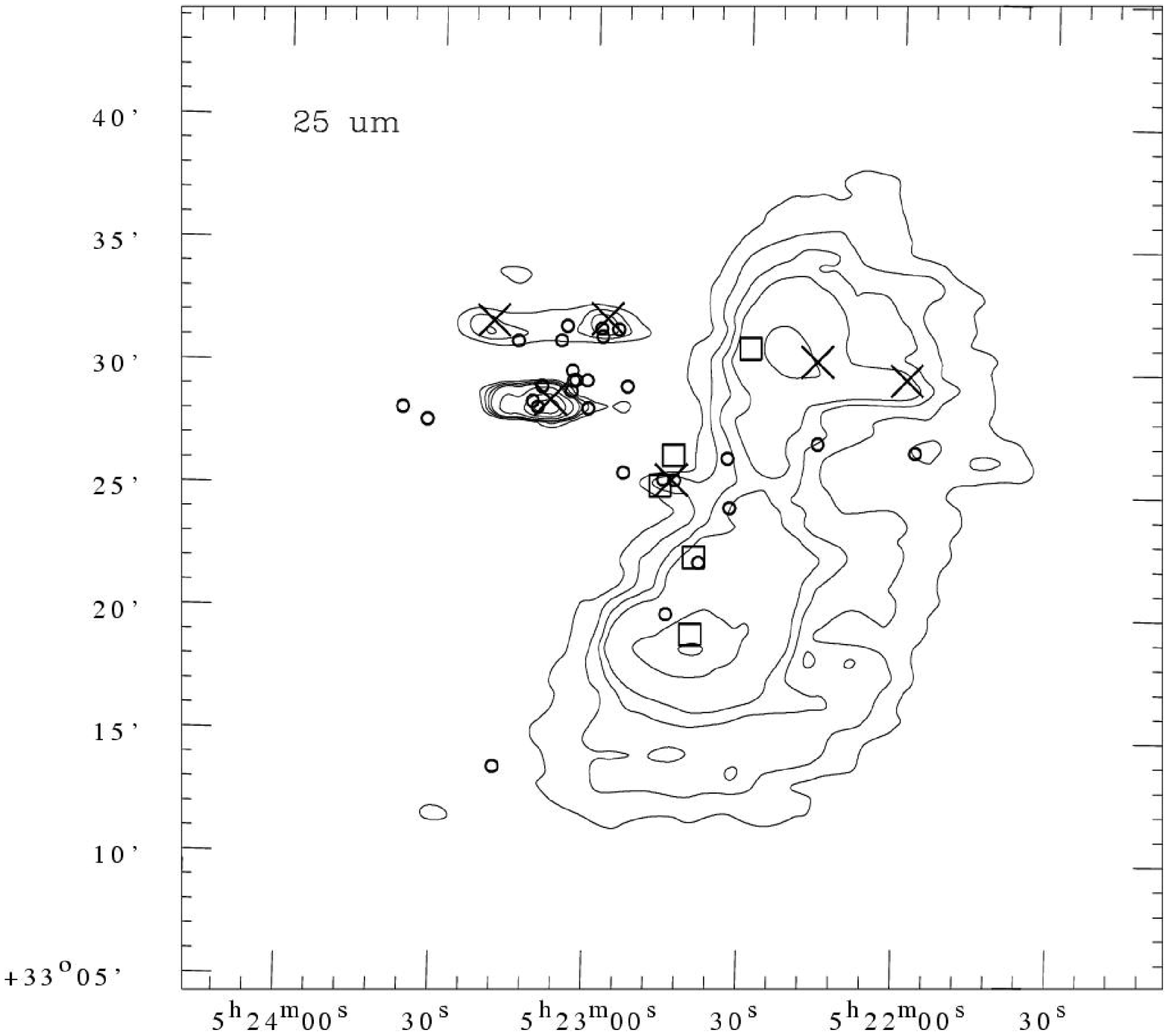}
}
\hbox{
\includegraphics[height=7cm,width=9cm]{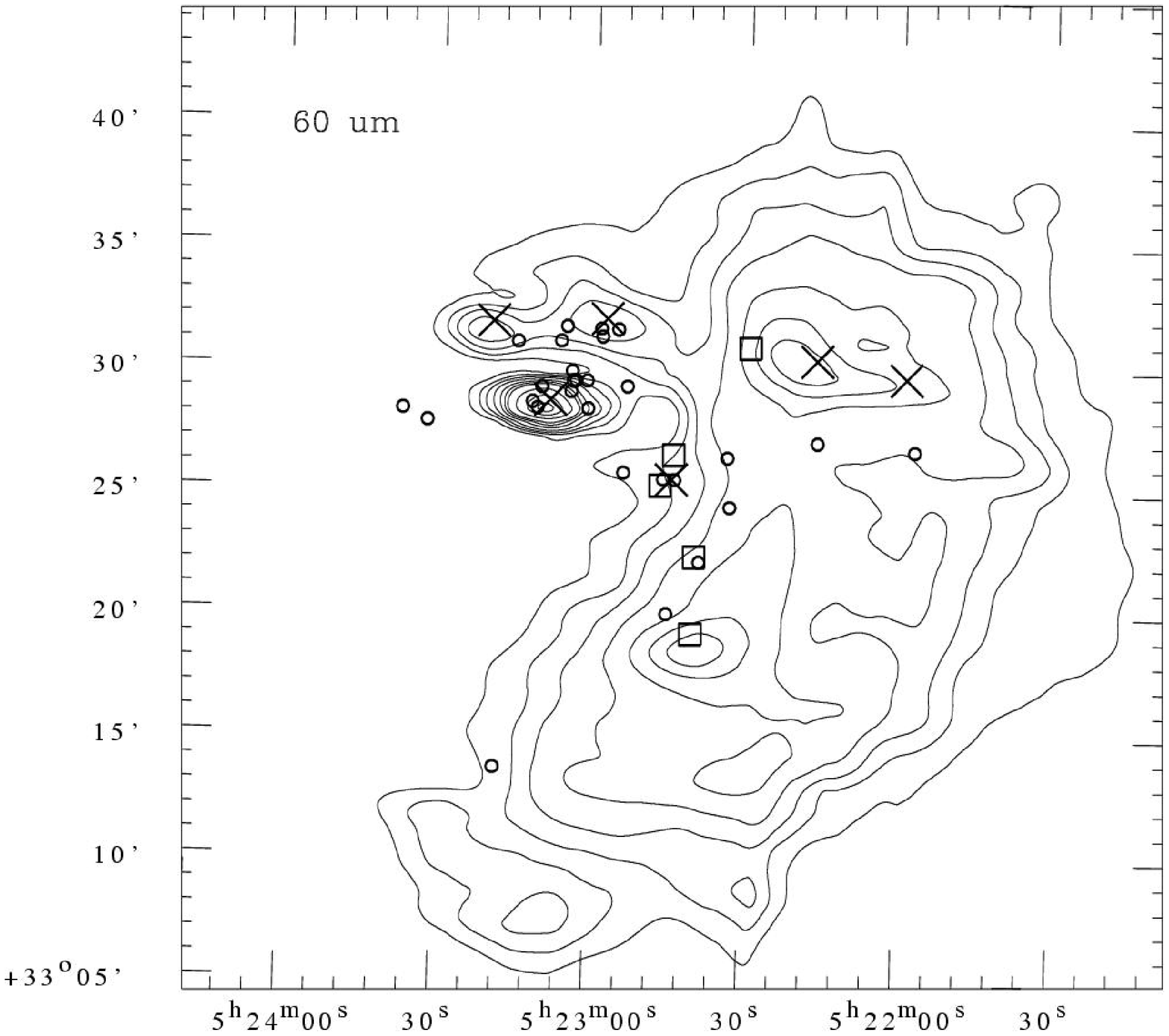}
\includegraphics[height=7cm,width=9cm]{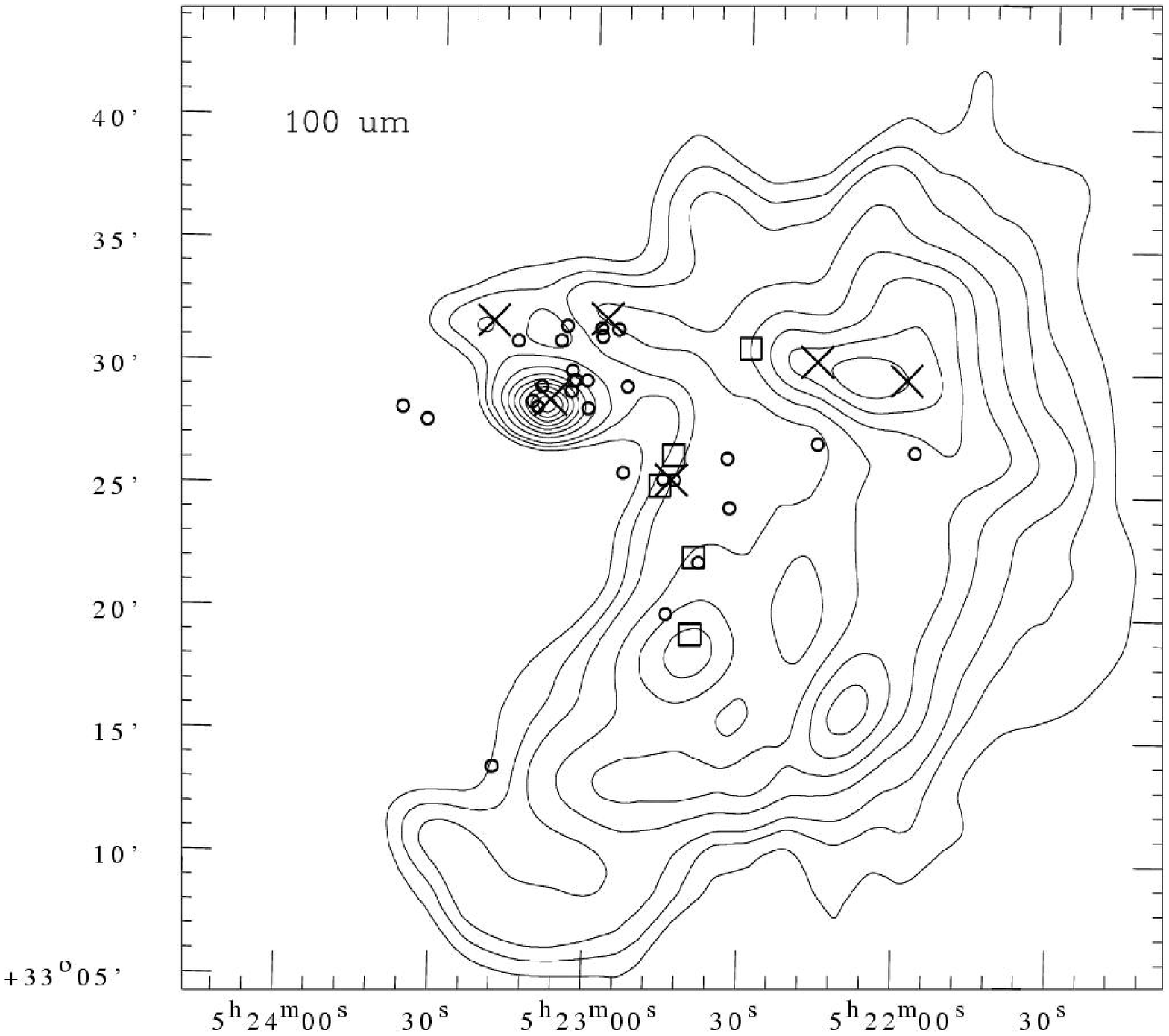}
}
\caption{The IRAS-HIRES intensity maps at $12~\mu$m $(top~ left)$, $25~ \mu$m $(top~ right)$,
$60~ \mu$m $(bottom~ left)$ and $100~ \mu$m $(bottom~ right)$. 
The contours are at 2, 3, 4, 5, 10, 20, 40, 60, 80\% of the peak value of 150 MJy/ster, 460 MJy/ster
at 12 and 25 $\mu$m respectively; 3, 5, 7, 10, 15, 20, 25, 40, 60, 80\% of the peak value of 
990 MJy/ster at 60 $\mu$m and 10, 15, 20, 30, 40, 50, 60, 70, 80, 90\% of the peak value of 
426 MJy/ster at 100 $\mu$m. The symbols are as same as in Fig. 22.}
\end{figure*}

\begin{figure*}
\centering
\hbox{
\includegraphics[height=7cm,width=9cm]{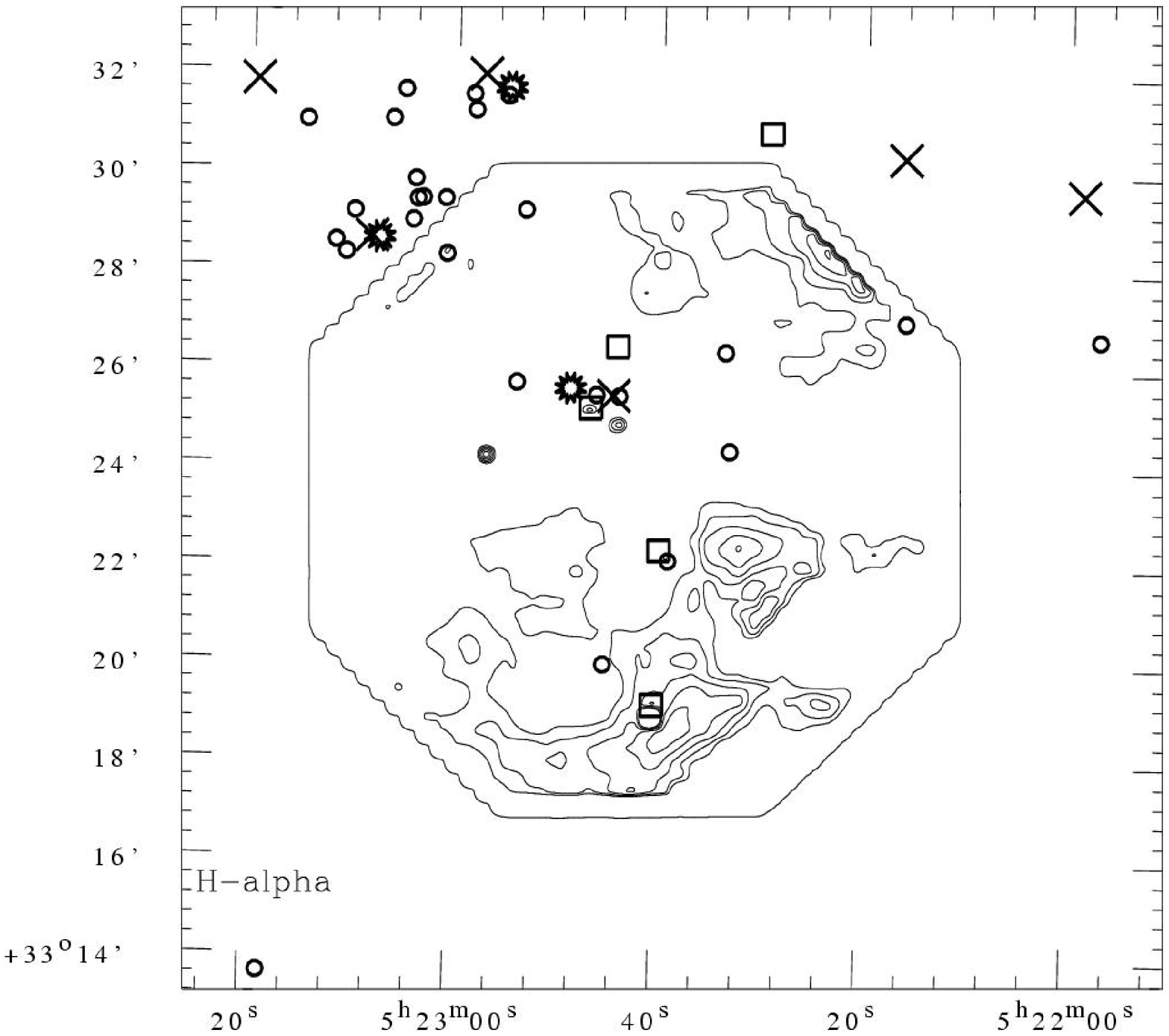}
\includegraphics[height=7cm,width=9cm]{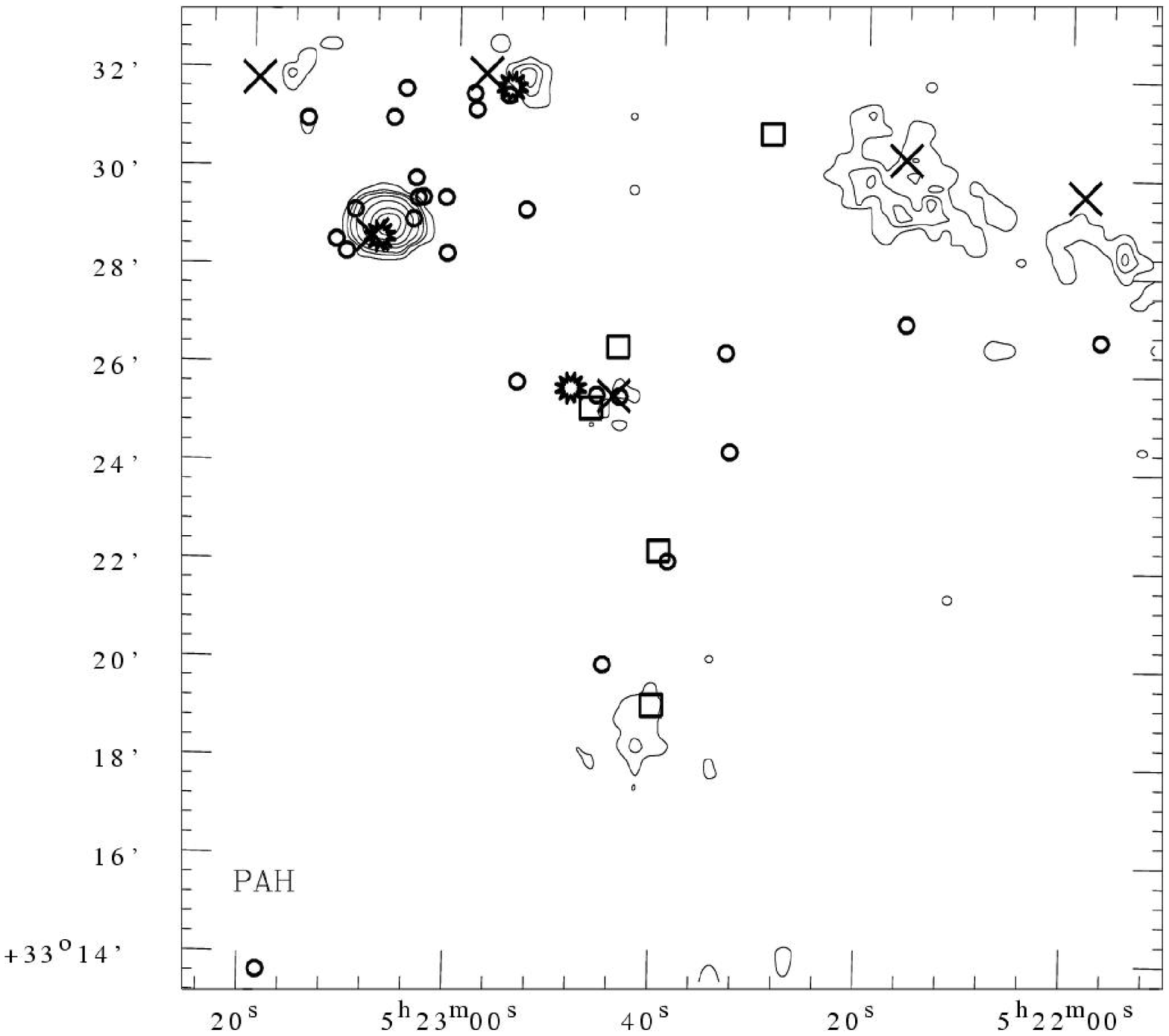}
}
\hbox{
\includegraphics[height=7cm,width=9cm]{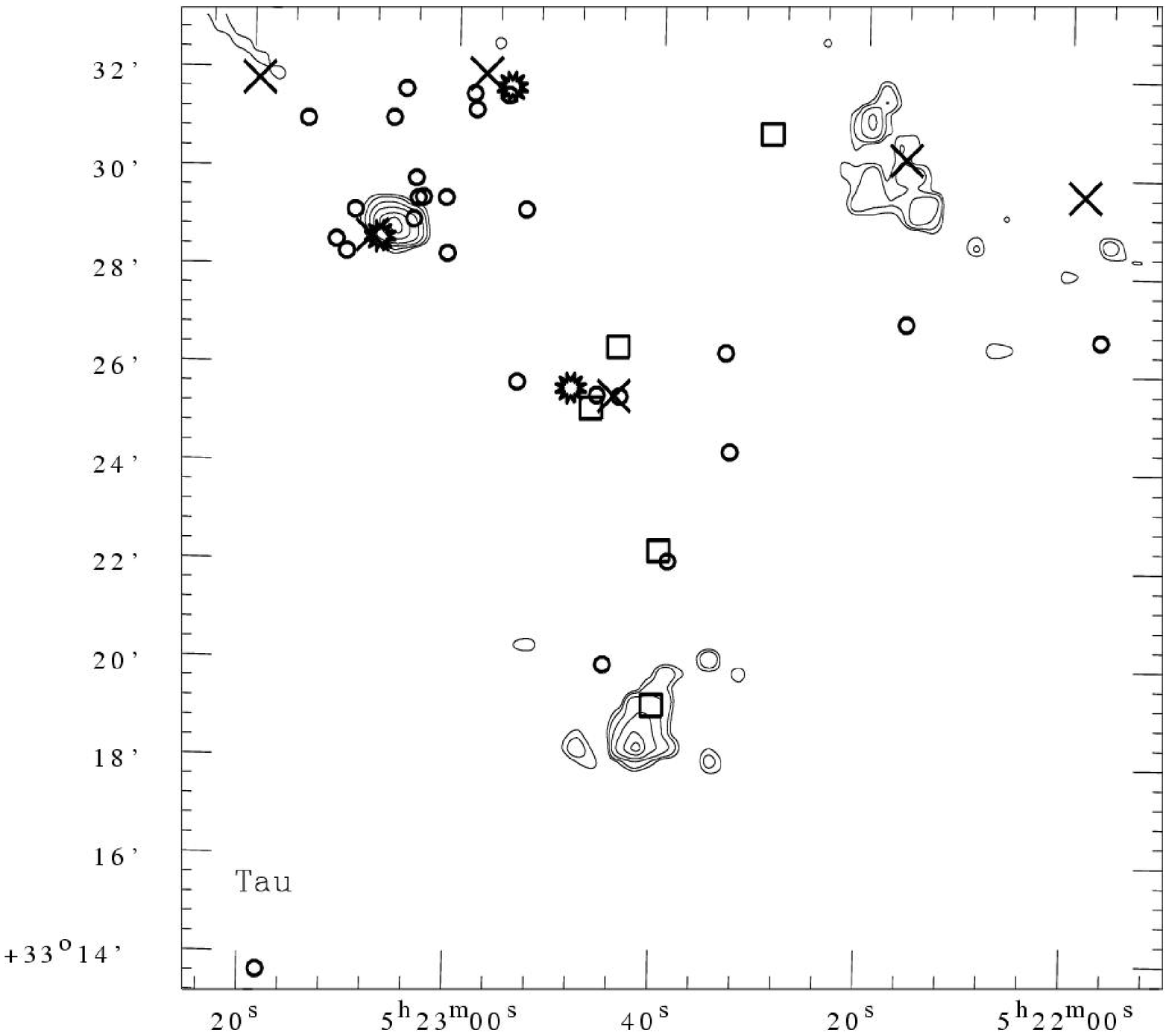}
\includegraphics[height=7cm,width=9cm]{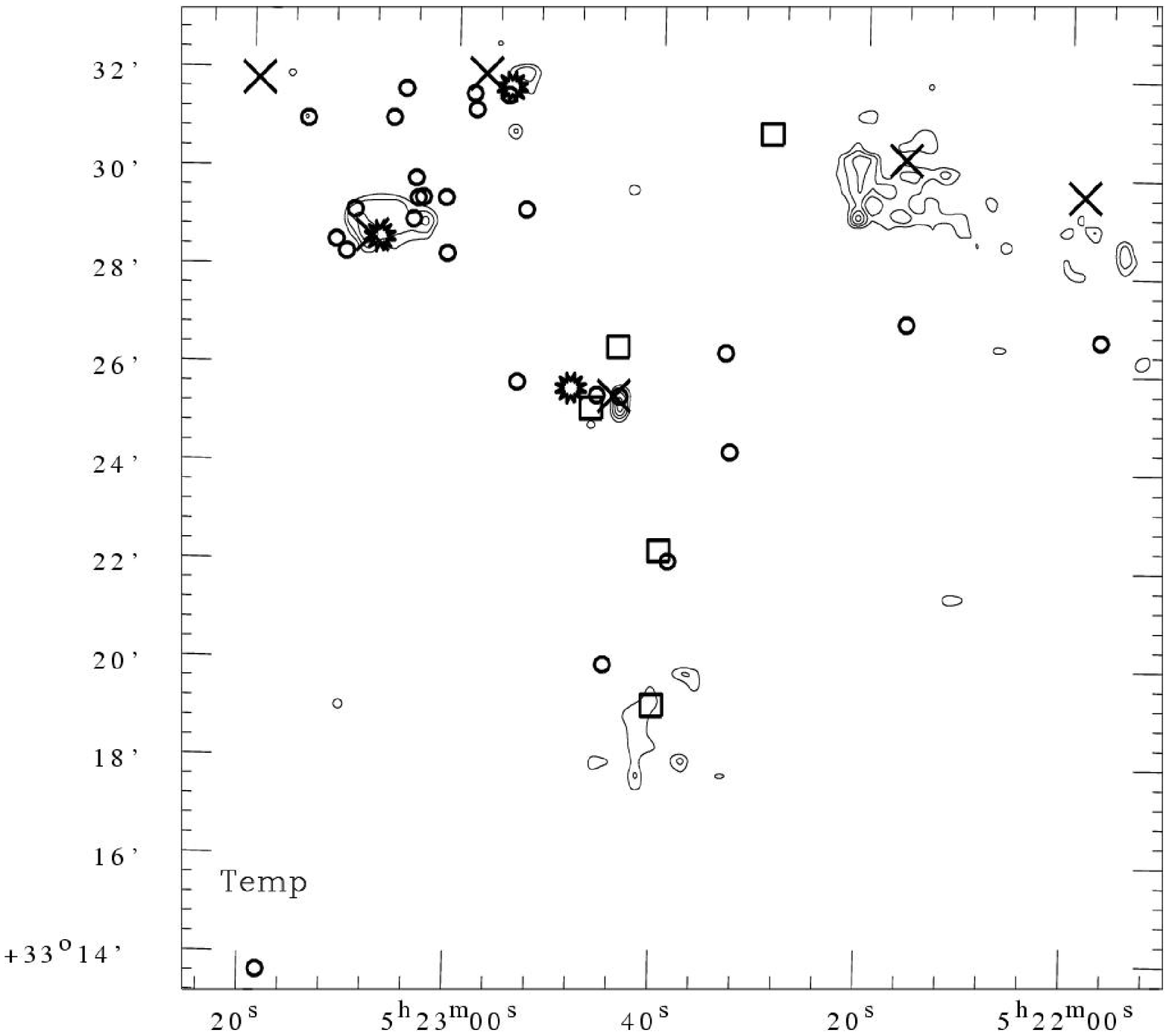}
}
\caption{$H\alpha$ intensity contours {\it(top left)}; hexagon represent the area of $H\alpha$ observations. 
Emission in UIBs {\it(top right)}, distribution of optical depth {($\tau_{10}$ at 10 $\mu$m)} 
{\it(bottom left)} and the distribution of dust temperature {\it (bottom right)}
around the cluster region. The contours are at 1, 3, 5, 10, 20, 40, 60, 80 \% 
of the peak value of $2.7\times10^{-5}$ $W m^{-2} Sr{^-1}$ for UIBs; 
3, 5, 10, 20, 40, 60, 80 \% of the peak value of  $8.5\times 10^{-5}$ for $\tau_{10}$
and 20, 30, 40, 50, 60, 70, 80, 90 \% of the peak value of 388 K
for dust temperature. The symbols are same as in Fig. 22. The star symbols are
the cluster center and positions of two nebulae, i.e., Sim 129 and Sim 130.
The abscissa and the ordinates are in the J2000.0 epoch.}
\end{figure*}

\begin{figure*}
\centering
\hbox{
\includegraphics[height=7cm,width=9cm]{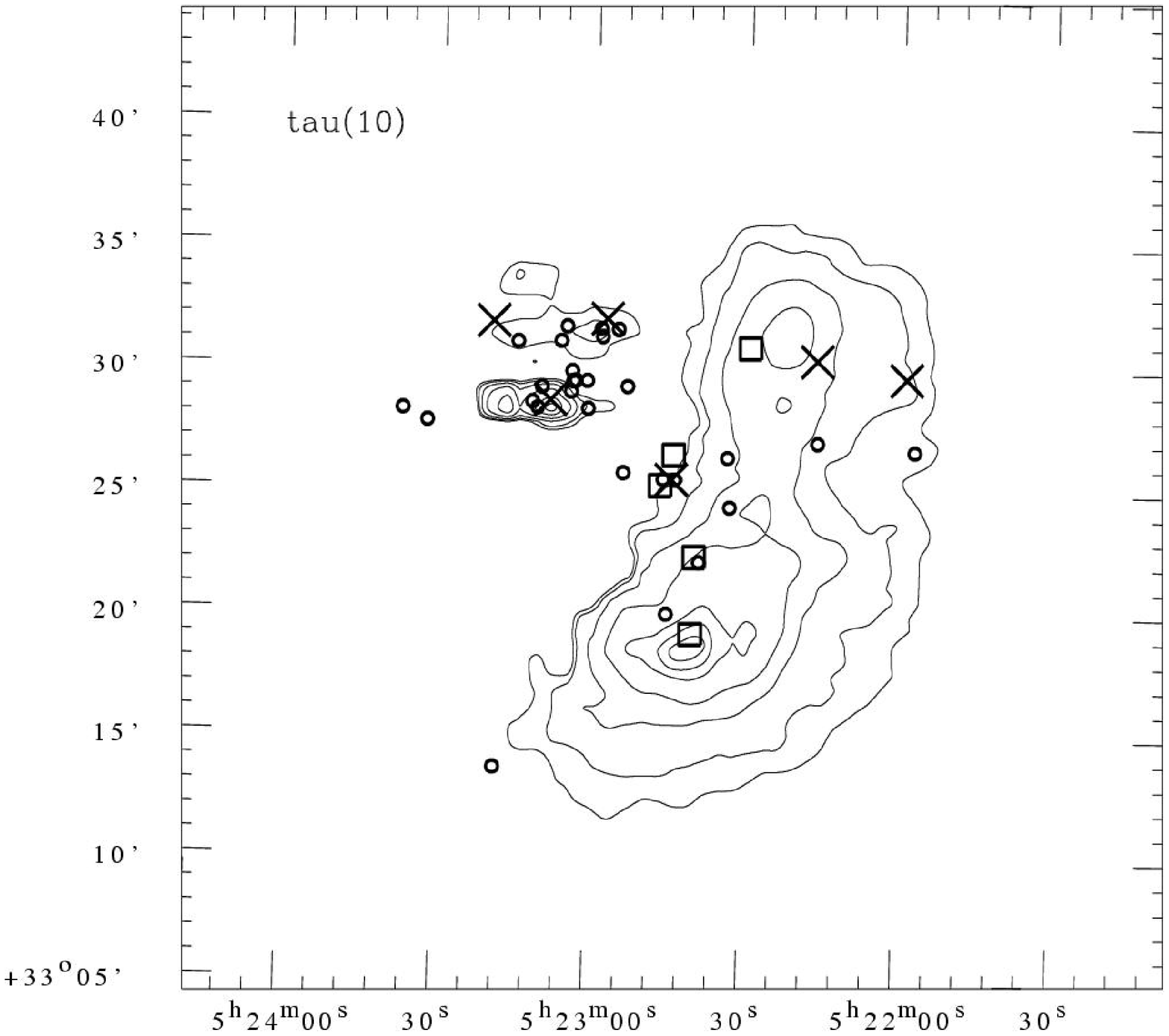}
\includegraphics[height=7cm,width=9cm]{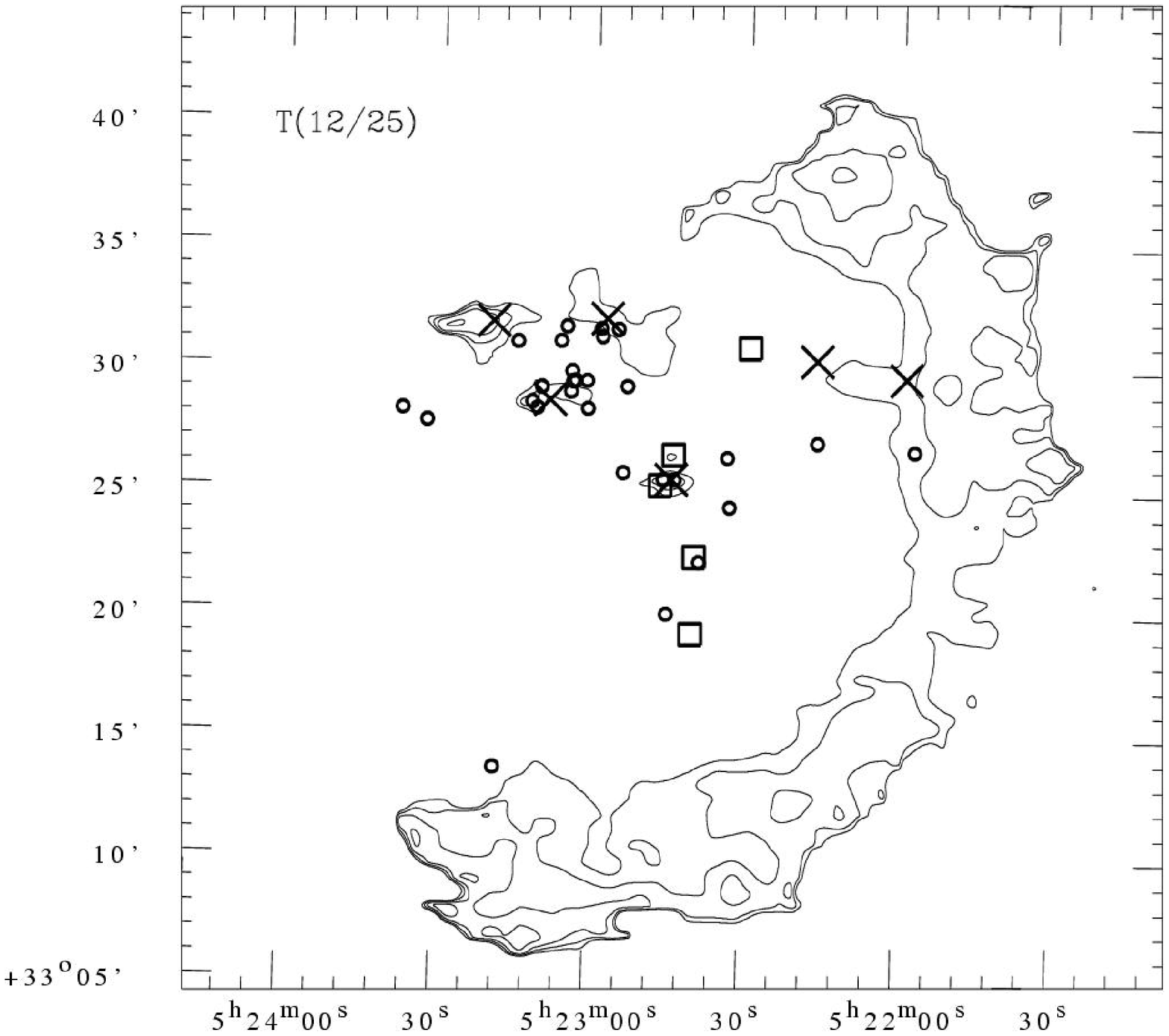}
}
\caption{{\it (Left)} The dust optical depth ($\tau_{10}$) distribution from the 
HIRES 12 and 25 $\mu$m maps, scaled to 10 $\mu$m by $\lambda ^{-1}$ emissivity law
and {\it (Right)} colour temperature (T (12/25)) (right)
maps of the region around the cluster. The optical depth contours are 3, 5, 10, 20, 40, 60, 80\%
of the peak value of $1.35 \times 10^{-5}$. The dust temperature contour levels are at
160, 171, 183, 197, 211, 226, 243, 260 and 279 K. The peak value is 300 K.}

\end{figure*}

The MSX A-band intensity map indicates presence of several discrete 
sources (including Sim 129 and 130) representing high-density clumps 
and region of further star formation. It is interesting to point out 
that 5 $H\alpha$, 2 NIR excess stars (see Fig. 22) and 2 PMS 
stars from Marco \& Negueruela (2002) lie on the rim of the cloud towards west of the cluster. This indicates, as suggested by Marco \& Negueruela (2002), this region is in the process of emerging from the parental cloud.

Fig. 24 shows the IRAS-HIRES intensity maps for the cluster region at $12~\mu$m (top~ left),  $25~ \mu$m (top~ right), $60~ \mu$m  (bottom~ left) and $100~ \mu$m (bottom~ right). The contour maps also show a bright arc shaped rim on the western side of the cluster, as seen in MSX as well as NVSS radio contour maps. Such a morphology implies that the HII region is ionization bounded towards the western side of the cluster. 

The cluster region shows emission in MIR (MSX A-band and IRAS $12~\mu$m) indicating presence of warm dust in the region but no radio emission.  The nebula Sim 129 is associated with the ionized region and also shows emission in  MIR and FIR 
(IRAS $12~\mu$m$ - 100~\mu$m), whereas the nebula Sim 130 does not show significant emission in FIR 
($60~\mu$m and $100~\mu$m). The clump to the NW of the cluster (hereafter clump A, $\alpha_{2000} = {05^h} {22^m.2}$, $\delta_{2000} = 33^{\circ} 33^{'}$), which harbors two IRAS sources, shows presence of ionized gas along with warm  dust (MSX A-band, IRAS $12~\mu$m and $25~\mu$m) as well as cold dust (IRAS $60~\mu$m$ - 100~\mu$m). This clump also shows $H\alpha$ emission (see Fig. 25). The above mentioned features indicate ongoing star formation activity in the clump. There is another clump towards south of the cluster (clump B, $\alpha_{2000} = {05^h} {22^m.6}$, $\delta_{2000} = 33^{\circ} 18^{'}.5$) which shows strong emission in FIR ($60~\mu$m$ - 100~\mu$m). This clump also shows emission in $H\alpha$ and an O7-type star is located in the clump B.

Distribution of YSOs and intensity maps of radio, NIR and FIR emissions 
in NGC 1893 region indicate four major clumps/condensations including 
Sim 129 and 130. YSOs distributed around  Sim 129 and 130 are found to 
have ages $\sim$ 1-3 Myr. The angular distance of Sim 129 and 130 from 
the cluster is about $\sim 5.^{'}6$ and $\sim 6.^{'}1$ respectively. 
Clump A (angular distance $\sim 10^{'}$) shows signature of ongoing star formation. 
A few NIR excess stars are visible towards the direction of the clump. 
As we have mentioned above, the region towards the clump A seems to be 
in process of emerging out from the parental cloud, so many more YSOs may still be 
embedded in the cloud.   

The contribution of UIBs due to PAHs to the mid-infrared emissions in the four MSX bands has been studied using the scheme developed by Ghosh \& Ojha (2002). The emission from each pixel is assumed to be a combination of two components, namely the thermal continuum from dust grains (gray body) and the emission from the UIB features in the MSX bands. The scheme assumes a dust emissivity of the power law form $\epsilon_\lambda \propto \lambda^{-1} $ and the total radiance due to the UIBs in band C is proportional to that in band A. A self
consistent non-linear chi-square minimization technique is used to estimate 
the total emission from the UIBs, the dust temperature, and the optical depth. 
The spatial distribution of the UIB emission, temperature map and optical 
depth ($\tau_{10}$ at 10 $\mu$m) contour map  with an angular resolution of $\sim 18^{\prime\prime}$ 
(for the MSX survey) extracted for the cluster region is shown in Fig. 25. 
Morphologically all are similar with the intensity peaks matching 
rather well with each other.  
This indicates the presence of high densities near the embedded sources.

We have used the IRAS-HIRES maps at 12 and 25 $\mu$m to generate the 
spatial distribution of dust colour temperature T(12/25) 
and optical depth at 25 $\mu$m ($\tau_{25}$) (cf. Ghosh et al. 1993). 
An emissivity law of $\epsilon_\lambda \propto \lambda^{-1} $ was
assumed to generate these maps. The dust colour temperature and optical depth
maps representing warmer dust component are presented in Fig. 26. The latter is 
shown in Fig. 26 ({\it left)}, which is scaled to 10 $\mu$m by a $\lambda^{-1}$ 
emissivity law to compare with the MSX $\tau_{10}$ map.
The distribution is centrally dense with optical depth peak and 
lower temperature at the center. 
The region is showing cometary appearance. Such a morphology can result if a 
massive star(s) is formed away from the center of a molecular cloud. This will 
cause the HII region to expand into an asymmetric density distribution and 
so becomes cometary (Israel 1978). From the orientation of the two emission nebulae 
and the arc shaped ring around the cluster center it is suggested that the central 
O type star(s) is (are) most likely responsible for the cometary morphology and 
the trigger of star formation. 

A comparison is also made between the $\tau_{10}$ maps generated from the
higher angular resolution MSX maps (Fig. 25: bottom left) and that based
on the IRAS HIRES maps at 12 and 25 $\mu$m (Fig. 26: left). 
The MSX $\tau_{10}$ map is brought to the same angular
resolution as of HIRES for the comparison. The peak optical depth is
8.54 $\times 10^{-5}$ for the map based on MSX. The corresponding value
from the IRAS-HIRES map is 1.36 $\times 10^{-5}$. These derived values
are in reasonable agreement considering that they are based on instruments
with very different angular resolutions. The difference in the peak
values of $\tau_{10}$ may be a result of the following effects: beam dilution,
clumpy interstellar medium and the contribution of UIB emission.
In particular in our study the contribution of the UIBs emission has been removed in 
generating the optical depth in the modeling of the MIR emission from 
the MSX bands, however the IRAS-HIRES maps represent the emission
from dust and UIB carriers (e.g. the emission from the UIB features
falls within the IRAS 12 $\mu$m band ($\sim$ 8-15 $\mu$m)).

The distribution of reddening (cf. Sec. 7.1) and close inspection of Fig. 22 indicate the presence of a region of low density of reddening material at the center of the cluster.
Similar trend has been noticed in a few H II regions (e.g. 30 Dor, Brandl et al., 1996 and NGC 3603, Pandey et al. 2000). A reasonable explanation for this lack of gas and dust in the central region may be a wind blown bubble by massive stars at the center of the cluster. 
Visual estimation of radius of the bubble from Fig. 22 yields a value of $\sim$ 13 - 14 arc minute corresponding to 12 - 13 pc at the distance of NGC 1893. 
A stellar wind of power $L_w$ $erg$ $s^{-1}$ from a star located in a uniform cloud of neutral hydrogen density, $n_0$ atoms $cm^{-3}$, will in time $`t$' yr produce a spherical shell of diameter $D(t)$ pc, expanding at $V(t)$ km $s^{-1}$, where

$L_w = 9.5 \times 10^{17} n_0~ V(t)^5~ t^2$

$D(t) = 3.5 \times 10^{-6}~V(t)~t$     (cf. Pandey et al. 2000) %, PASJ,52, 847) 

Assuming that O7 stars at the center of the cluster are injecting a kinetic energy $L_W$ $\sim$ $10^{36} ~erg$ $s^{-1}$, with the gas density $10^3 - 10^4 ~cm^{-3}$ and age of O-type stars $\sim$ 4 Myr, the radius of the bubble corresponds to $\sim$ 10 - 16 pc.

\section{Summary and Conclusion}

On the basis of a comprehensive multi-wavelength study of the star forming region NGC 1893 we have made an attempt to study the effects of massive stars on low mass star formation. Deep optical $UBVRI$ and narrow band $H\alpha$ photometric data, slit-less spectroscopy along with archival data from the surveys such as 2MASS, MSX, IRAS and NVSS are used to understand the global scenario of star formation in and near the cluster region.  

Reddening ($E(B-V)$) in the direction of cluster is found to be varying between 0.40 to 0.60 mag. The post-main-sequence age and distance of the cluster are found to be $\sim 4$ Myr and  3.25 $\pm 0.20$  kpc respectively. Using the NIR two colour diagram and excess $H\alpha$ emission  we identified candidate YSOs which are aligned from the cluster to the direction of nebula  Sim 129. The O-type stars at the center of the cluster may be responsible for the trigger of star formation in the region. The morphology of the cluster seems to be influenced by the star formation in the region. We find a population of PMS YSOs in the cluster region having mass $\sim 1 - 3.5 M_\odot$. The position of the YSOs on the CMDs indicates that the majority of these stars have age between $\sim$ 1 Myr to 5 Myr indicating a possibility of non-coeval star formation in the cluster. Two probable HBe stars in the vicinity of central O-type stars further indicate an prolonged star formation in the center of the cluster region.

The K-band luminosity function for the cluster is found to be $0.34\pm0.08$ which is 
consistent with the average value ($\sim$ 0.4) obtained for young star 
clusters (Lada et al. 1991; Lada \& Lada 1995; Lada \& Lada 2003). The slope of the initial mass function $`\Gamma$' 
for PMS stars (mass range $0.6 < M/M_\odot \le 2.0 $) is found to be $-0.88\pm0.09$, 
which is shallower than the value ($-1.71\pm0.20$) obtained for MS stars having 
mass range $2.5 < M/M_\odot \le 17.7$. However for the entire mass range 
($0.6 < M/M_\odot \le 17.7 $) the $`\Gamma$' comes out to be $-1.27\pm0.08$. 
The effect of mass segregation can be seen on the MS stars. The estimated dynamical 
evolution time is found to be greater than the age of the
cluster, therefore the observed mass segregation in the cluster may be the imprint
of the star formation process.

\section{Acknowledgements}

Authors are thankful to the anonymous referee for useful comments. 
The observations reported in this paper were obtained using the 2 meter HCT at IAO, Hanle, the high altitude station of Indian Institute of Astrophysics, Bangalore. We thank the staff at IAO, Hanle and its remote control station at CREST, Hosakote for their assistance during observations. This publication makes use of data from the Two Micron All Sky Survey, which is a joint project of the University of Massachusetts and the Infrared Processing and Analysis Center/California Institute of Technology, funded by the National Aeronautics and Space Administration and the National Science Foundation. AKP is thankful to the National Central University, Taiwan for the financial support during his visit to NCU. The support given by the DST (India) is also thankfully acknowledged.

\bsp

\label{lastpage}

\end{document}